\documentclass[fleqn,usenatbib,useAMS]{mnras}

\usepackage{newtxtext,newtxmath}
\usepackage[T1]{fontenc}
\usepackage{graphicx}	
\usepackage{amsmath}	
\usepackage{multicol}   
\usepackage{multirow}
\usepackage{bm}		    
\usepackage[utf8]{inputenc}
\usepackage[english]{babel}
\usepackage{algorithm}
\usepackage[toc,page]{appendix}
\usepackage{hyperref}
\usepackage{xspace}
\usepackage{url}
\usepackage[table]{xcolor}
\usepackage{longtable}
\usepackage{listings}
\usepackage{placeins}
\usepackage{array}
\usepackage{tikz}

\newcommand{\fwhm}{\text{FWHM}}

\newcommand{\logRhk}{$\log R'_{HK}$\xspace}
\newcommand{\tav}{\ensuremath{\delta t_{\rm av}}\xspace}
\newcommand{\tspan}{\ensuremath{t_{\rm span}}\xspace}

\definecolor{pythonblue}{RGB}{31,119,180}
\definecolor{pythongreen}{RGB}{44,160,44}
\definecolor{pythonred}{RGB}{214,39,40}

\title[Modelling stellar activity with GPRNs]%
    {Modelling stellar activity with Gaussian process regression networks}

\author[J. D. Camacho et al.]{
    J. D. Camacho,$^{1,2}$\thanks{E-mail: joao.camacho@astro.up.pt}
    J. P. Faria,$^{1,2}$
    P. T. P. Viana,$^{1,2}$
\\
    $^{1}$~Instituto de Astrofísica e Ciências do Espaço, Universidade do Porto, 
    Rua das Estrelas, 4150-762, Porto, Portugal \\
    $^{2}$~Departamento\,de\,Física\,e\,Astronomia,\,Faculdade\,de\,Ciências,\,%
    Universidade\,do\,Porto,\,Rua\,do\,Campo\,Alegre,\,4169-007,\,Porto,\,Portugal \\
}

\date{Accepted XXX. Received YYY; in original form ZZZ}

\pubyear{2022}

\begin{document}
\label{firstpage}
\pagerange{\pageref{firstpage}--\pageref{lastpage}}
\maketitle

\begin{abstract}
    Stellar photospheric activity is known to limit the detection and
    characterisation of extra-solar planets. In particular, the study of
    Earth-like planets around Sun-like stars requires data analysis methods that
    can accurately model the stellar activity phenomena affecting radial
    velocity (RV) measurements. Gaussian Process Regression Networks (GPRNs)
    offer a principled approach to the analysis of simultaneous time-series,
    combining the structural properties of Bayesian neural networks with the
    non-parametric flexibility of Gaussian Processes. Using HARPS-N solar
    spectroscopic observations encompassing three years, we demonstrate that
    this framework is capable of jointly modelling RV data and traditional
    stellar activity indicators. Although we consider only the simplest GPRN
    configuration, we are able to describe the behaviour of solar RV data at
    least as accurately as previously published methods. We confirm the
    correlation between the RV and stellar activity time series reaches a
    maximum at separations of a few days, and find evidence of non-stationary
    behaviour in the time series, associated with an approaching solar activity
    minimum.
\end{abstract}


\section{Introduction}
\label{ch:Introduction}

The detection of extra-solar planets, henceforth exoplanets, as a research field
in Astronomy started to develop from the discovery of 51 Pegasi \textit{b} by
\cite{Mayor1995}. After the identification of this hot-Jupiter, many more
followed with over 5000 exoplanets found so far. The two most successful
techniques to detect exoplanets are the radial velocity (RV) and transit
methods. The former relies on the fact that both planet and star orbit a common
centre of mass, which generates a Doppler shift in the spectrum of the star that
can be detected by high precision spectrographs.

Unfortunately, the use of RV data to detect and characterise exoplanets is not
without its challenges. In particular, stellar magnetic activity is known to
impact the RV measurements, imitating and/or hiding planetary signals
\citep[e.g.][]{Figueira2010_BD201790, Santos2010_StellarMagCycles,
Santos2014_HD41248}. Several activity indicators are commonly used to
disentangle the activity and planetary signals present in the RVs, aiding the
identification of periodic signals as stellar or planetary in origin
\citep[e.g.][]{Figueira2013_Indicators}. More recently, the use of Gaussian
processes \citep[GP, ][]{Rasmussen2006_GPs} has proved to be a successful tool to
model RV data. While the GP covariance function describes the behaviour of the
stellar RV contribution, the mean function is assumed to be generated by
whatever planetary signals may be present \citep[e.g.][]{Haywood2014_Corot7,
Faria2016_Corot7, Cloutier2019_K218}.

A new generation of spectrographs, e.g. EXPRES \citep{Petersburg2020_EXPRES}
and ESPRESSO \citep{Pepe2021_ESPRESSO_DRS}, is providing RV measurements precise
enough to enable the detection of Earth-like planets orbiting in the habitable
zone of Sun-like stars. However, such planets cause a RV signal of the order of
10 cm/s, easily overwhelmed by the impact of stellar activity. Advanced
statistical methods capable of characterising such feeble planetary signals are
thus required. In this paper, we use a Gaussian process regression network
(GPRN) to analyse solar RV data together with activity indicators, such as the
full width at half maximum (\fwhm) and the \logRhk index, and show that this
framework can successfully describe the impact of stellar activity on RV data.

We start by introducing, in section~\ref{ch:stellarActivityChallenges}, the
challenges that stellar activity brings to exoplanet detection with RVs and with
the definition of the main activity indicators used to identify stellar activity
signals. In section~\ref{ch:GPRN}, we present the GPRN framework, followed by
some tests with simulated data in section~\ref{ch:AnalysisSamples}. We describe
the HARPS-N solar observations and interpret the results from applying the GPRN
in section~\ref{ch:Sun}. Finally, section~\ref{ch:DiscussionConclusions}
presents a discussion and our conclusions.

\section{The stellar activity RV imprint}
\label{ch:stellarActivityChallenges} 

Stellar photospheric phenomena span a wide range of characteristic timescales,
thus requiring different strategies in order for its impact to be addressed.
Namely,

\begin{itemize}
    \item[-] on the timescale of a few minutes, pressure waves (p-modes) lead to
    the contraction and dilation of the stellar external envelope, inducing RV
    signals of the order of 1 m/s \citep{SchrijverZwaan2000_pModes,
    Bazot2007_oscillations};
    \item[-] within a timescale of minutes to hours, granulation and
    super-granulation phenomena caused by convection in the external layers of
    solar-type stars induce signals with RV amplitudes also in the range of 1
    m/s \citep{Kjeldsen2008_oscillations, Meunier2019_SuperGranulation};
    \item[-] due to the variation of the stellar radius, gravitational
    red-shift induces variations in the RV measurements of the order of a few
    cm/s \citep{Cegla2012_GravRedshift};  
 
    \item[-] magnetic activity related to spots, faculae, and plages induces
    quasi-periodic signals with amplitudes between 1 and 50 m/s, depending on
    the activity level and age of the stars \citep{Saar1997,
    Santos2010_StellarMagCycles, Lagrange2010_ColdSpots}, and characteristic
    periodicities ranging from the stellar rotation period up to the magnetic
    cycle, i.e. from days to years \citep{Saar1997, Baliunas1995_magCycles}.

\end{itemize}

The RV impact of pressure waves and the different types of granulation can be
addressed, at least in part, using specific observational strategies
\citep[e.g.][]{Dumusque2011a}. But the activity signals induced by active
regions need to be modelled, since they encompass the most interesting range of
orbital periods from an habitability point of view, and are known to hide or
mimic planetary signals in RV measurements \citep[e.g.][]{Robertson2014_G581,
Santos2014_HD41248, Guinan2016_KapteynStar}. 

However, the joint identification and characterisation of the complex RV signals
induced by stellar magnetic activity and those due to orbiting planets is often
challenging. One way to tackle this issue is by using the information contained
in line-profile or chromospheric activity indicators, which by construction
should not be affected by any orbiting planets. In this work, we focus on three
such indicators, the FWHM, the BIS, and the \logRhk index \citep[see,
e.g.,][]{Figueira2013_Indicators,Gomesdasilva2012_Activity}. These are estimated
jointly with the RVs through analysis of the cross-correlation function (CCF)
and the Calcium H and K lines.

Nevertheless, our framework is general enough to accommodate any number of
activity proxies which are observed simultaneously with the RVs. For example,
stellar magnetic activity also affects photometric observations and
high-precision photometry has been used as a proxy when characterising planet
signals in RV data \citep[e.g.][]{Haywood2014_Corot7,
Kosiarek2020_photometryAsProxy}. Yet, as it is currently implemented (see
section \ref{ch:GPRN}), the GPRN requires that all time-series have equal
timestamps, which precludes the use of non-simultaneous photometric
observations, even if they are contemporaneous. This limitation may be addressed
in future work.

\section{Gaussian process regression networks}
\label{ch:GPRN}

In a seminal paper, \citet{Rajpaul2015_GPFramework} proposed a framework to
jointly model RVs and several activity indicators. The key insight was to extend
the FF' method from \citet{Aigrain2012_FFMethod}, which relates the activity
signal expected in the RVs to the relative drop in flux caused by point-like
spots (denoted F) and its first time derivative. From this premise,
\citet{Rajpaul2015_GPFramework} modelled F with a latent Gaussian Process (GP)
and derived the necessary expressions for a multi-output model based on linear
combinations of F and F'. They used the \logRhk and the BIS because of their
sensitivity to the spot coverage on the star. 

More recently, \citet{Barragan2021_pyaneti2} extended this approach to a more
general set of activity indicators and presented an efficient implementation
within the \texttt{pyaneti} package \citep{Barragan2019_pyaneti1}. The framework
has also been extended by \citet{Jones2017_GPFramework}, who included terms
proportional to the second time derivative of the latent GP and by
\citet{Gilbertson2020_JonesFramework}, who developed the \texttt{GLOM} model
allowing for the use of any covariance kernel. This methodology has proven
useful in disentangling planetary and stellar induced signals in RV data with
different levels of activity~\citep[e.g][]{Barragan2019_K2100b,
Mayo2019_Kepler538b} .

While these approaches are an improvement with respect the
traditional GP regression framework that models only RVs, they are limited by
their inability to tackle the non-stationary behaviour of stellar magnetic
activity as a function of time \citep[e.g.][]{Demin2018_NonStatSolarActivity},
i.e. they are incapable of addressing processes whose characteristics vary
across the input space \citep{Plagemann2008_nonStationaryGPs}. The development
of a framework that combines the information present in activity indicators and
is also capable of tackling non-stationary should thus help the identification
of the RV component due to stellar activity and improve planet detection and
characterisation. This was the motivation behind the implementation of a
Gaussian process regression network (GPRN) framework as a means for the joint
analysis of RV data and activity indicators.

\subsection{Gaussian processes}
\label{ch:GPs}

We first review the basic concepts behind a more traditional GP, which will also
be useful for the presentation of the GPRN framework.. In brief, a GP is a
generalisation of the multivariate Gaussian distribution, characterised by a
mean function $m(t)$ and a covariance function $k(t,t')$, more commonly known as
the \textit{kernel}. More formally, a GP is a stochastic process with Gaussian 
marginal distributions, thus defining a distribution over functions, 
i.e. each draw from a GP represents a function~\citep{Quadrianto2010_GPs}.

A GP is sufficiently flexible to model the quasi-periodic covariance structure
induced by stellar activity in RV measurements, by means of its kernel, while
also being able to take into account Keplerian contributions to the RVs through
its mean function \citep[e.g.][]{Haywood2014_Corot7, Faria2016_Corot7,
Cloutier2019_K218}. The most commonly used GP kernel in RV analysis is the
quasi-periodic (QP) kernel, obtained from the multiplication of a periodic and a
squared-exponential covariance function:
\begin{equation}
\label{eq:QPkernel}
    k_{\mathcal{QP}}\left(t,t'\right) = 
        \eta^{2}_{1} \exp \left[ 
            - \frac{\left(t-t'\right)^2}{2\eta^{2}_{2}} 
            - \frac{2}{\eta^{2}_{4}} \sin^2\left(\frac{\pi (t-t')}{\eta_3} \right) 
        \right].
\end{equation}
The kernel hyper-parameters $\bm{\eta}$ can have a physical interpretation. While
$\eta_1$ relates to the strength of the correlation between RVs at different
time separations, $\eta_2$ and $\eta_3$ reflect the evolution timescale of the
active regions at the stellar surface and the stellar rotation period,
respectively. Lastly, $\eta_4$ controls how sinusoidal the GP latent functions
are or, in other words, their timescale of variation relative to $\eta_3$. For
smaller values of $\eta_4$, the functions will show more short-scale structure
within one stellar rotation period \citep[e.g.][]{Rajpaul2017_thesis}. 

With the mean and covariance functions defined and parameterised, the next step
is to condition the GP on the observed data. We start by writing the log
marginal likelihood of a set of observations $\bm{y}$ under the GP,
\begin{equation}
\label{eq:logLikelihood}
\log \mathcal{L} = - \frac{1}{2} \bm{r}^T K^{-1} \bm{r} 
                   - \frac{1}{2}\log |K| 
                   - \frac{n}{2} \log\left(2\pi \right),
\end{equation}
\noindent where $\bm{r}$ is the vector of residuals after subtraction of the
mean function, $K$ the covariance matrix of the GP, and $n$ the number of
measurements. The three terms in Equation~\eqref{eq:logLikelihood} can be
interpreted as a measure of the goodness-of-fit, a penalisation for the
complexity of the covariance function, and a normalisation constant,
respectively \citep[see][]{Rasmussen2006_GPs}.

The marginal likelihood can be maximised to obtain a set of `best'
hyper-parameters compatible with the data. Alternatively, one can place prior
distributions on each of the hyper-parameters (of the mean and covariance
functions) and sample from their posterior distributions using, for example,
MCMC. For this work, we explored several publicly available MCMC packages such
as \texttt{emcee} \citep{ForemanMackey2019_emcee}, \texttt{dynesty}
\citep{Speagle2020_dynesty}, and \texttt{zeus} \citep{Karamanis2021_zeus},
finally settling on \texttt{emcee}.

\subsection{A GPRN model for stellar activity}

Jointly modelling RVs and activity indicators can be framed as a multi-output
regression problem. Generalising GPs for multi-output problems is not a trivial
task, since the statistical dependence across the different datasets (or
data \textit{channels}) must be explicitly modelled. Here we explore GPRNs
\citep{Wilson2012_GPRN} as a set of promising Bayesian models for multi-output
regression which exploit the structural properties of neural networks and the
flexibility of non-parametric function learning offered by GPs. As we will see,
GPRNs can capture input-dependent, highly non-linear correlations between the
outputs, provide heavy-tailed predictive distributions, and can resist
over-fitting.

Following \citet{Wilson2012_GPRN}, we write the GPRN model for a set of $P$
outputs $\bm{y}(t)$ as
\begin{equation}
\label{eq:GPRN}
\bm{y}(t) = \bm{W}(t) \left[ \bm{f}(t) + \sigma_f \bm{\epsilon}(t) \right] %
            + \sigma_y \bm{z}(t)
\end{equation}
\noindent
where we introduced a small set of $Q$ latent functions (called \textit{nodes})
$\bm{f}(t)$ and a $P \times Q$ projection matrix $\bm{W}(t)$ of \textit{weight}
functions. Both the nodes and the weights follow independent GP priors 
\begin{align}
\label{eq:NodesWeights}
    f_j (t)    \sim & ~ \mathcal{GP}(0, k_{f_j})~\text{for}~j = 1, ..., Q, \\
    W_{ij} (t) \sim & ~ \mathcal{GP}(0, k_{w_{ij}})~\text{for}~i = 1, ..., P~\text{and}~j = 1, ..., Q
\end{align}
\noindent
with zero means and kernels $k_{f_j}$ and $k_{w_{ij}}$, respectively. In
equation \eqref{eq:GPRN}, both $\bm{\epsilon}(t)$ and $\bm{z}(t)$ are
independent, white noise processes sampled from the standard normal
distribution. We make the simplifying assumption of setting $\sigma_f=0$, i.e.
we assume a priori that the white noise component of the model is stationary,
akin to a so-called jitter term. We associate one such term, characterised by
its standard deviation $\sigma_i$, to each dataset $i$. However, we also have to
take into account the independent measurement uncertainties that we know affect
the data. Each measurement $n$ in dataset $i$ has an associated uncertainty
described by a normal distribution with standard deviation $\sigma_{i,n}$.
Therefore
\begin{equation}
\label{eq:sigmaCalculation}
    \sigma_y \rightarrow \sigma_{y_{i}}=\sqrt{ \sigma_{i}^2 + \sigma_{i,n}^2 }
\end{equation}
where each $\sigma_{y_{i}}$ depends implicitly on $n$.

The overall structure of the GPRN creates a non-stationary linear combination of
independent GPs. The correlations between the various variables are modelled by
the latent functions $\bm{f}(t)$ \citep{Nguyen2013b_GPRN}. The final amplitudes
are given by projection through the weight functions $\bm{W}(t)$, making the
correlations input dependent and resulting in non-stationary outputs
\citep{Heinonen2015_nonStationaryGPs}.

In the simplest case the GPRN has two latent variables, one node and one weight,
causing $a$ $priori$ the output to have a marginal distribution defined by the
product of two independent Gaussian distributions. This product is not in
general a Gaussian distribution but instead follows a re-normalised Bessel
distribution of the second kind and order zero \citep[e.g.][]{Arfken2013_Bessel,
Cui2016_Bessel}, which has longer tails than a Gaussian. This will occur
regardless of the type of covariance function used for the node and the weight
GPs. Thus, the GPRN posterior predictive distribution has the flexibility 
to be heavier-tailed than the (Gaussian) predictive of a GP, making it more robust 
with respect to the possible presence of outliers. This also implies that some 
popular goodness-of-fit metrics, like the root mean square (rms) residuals, should 
be applied with care to GPRN results, especially when comparing GPRN residuals to
those from a GP.

The covariance functions for the latent GPs, $\bm{f}(t)$ and $\bm{W}(t)$, can
take any form (periodic, quasi-periodic, squared exponential) and they can be
shared among all nodes or all weights, or individually defined. This allows for
a model with an overall covariance kernel that continuously shifts between
regions with completely different covariance structures, thus accommodating
non-stationary correlations in the outputs \citep{Wilson2014_Thesis,li2020_gprn}. 
This suggests a GPRN can be used to model the complex, non-periodic 
signals with different decaying timescales and non-stationary amplitudes that 
are often the result of stellar activity.

\begin{figure}
    \begin{center}
        \begin{tikzpicture}
            \draw [black] (0,0) rectangle (0.75,0.75);
    \node at (0.375,0.375) {t};
    \draw [->] (0.75,0.375) -- (2.5,2.25);
    \draw [->] (0.75,0.375) -- (2.5,1);
    \draw [->] (0.75,0.375) -- (2.5,-1.5);

    \draw [pythonred, fill=pythonred] (3.0,2.25) circle [radius=0.5];
    \node [black] at (3.0,2.25) {$f_{1}$};
    \draw [pythonred, ->] (3.5,2.25) -- (5.25, 2.25);
    \draw [pythonred, ->] (3.5,2.25) -- (5.25, 1.0);
    \draw [pythonred, ->] (3.5,2.25) -- (5.25, -1.5);
    \draw [pythongreen, fill=pythongreen] (3.0,1.0) circle [radius=0.5];
    \node [black] at (3.0,1.0) {$f_{2}$};
    \draw [pythongreen, ->] (3.5,1.0) -- (5.25, 2.25);
    \draw [pythongreen, ->] (3.5,1.0) -- (5.25, 1.0);
    \draw [pythongreen, ->] (3.5,1.0) -- (5.25, -1.5);
    \draw [white] (3.0,-0.6) circle [radius=0.5];
    \node at (3.0, -0.2) {$(\cdots)$};
    \node at (5.75, -0.2) {$(\cdots)$};
    \draw [pythonblue, fill=pythonblue] (3.0,-1.5) circle [radius=0.5];
    \node [black] at (3.0,-1.5) {$f_{Q}$};
    \draw [pythonblue, ->] (3.5,-1.5) -- (5.25, 2.25);
    \draw [pythonblue, ->] (3.5,-1.5) -- (5.25, 1.0);
    \draw [pythonblue, ->] (3.5,-1.5) -- (5.25, -1.5);

    \draw [black] (5.25,1.75) rectangle (6.4, 2.75);
    \node at (5.8, 2.25) {RV};
    \draw [black] (5.25,0.50) rectangle (6.4, 1.5);
    \node at (5.8,1.0) {AI$_1$};
    \draw [black] (5.25,-2.0) rectangle (6.4, -1.0);
    \node at (5.8,-1.5) {AI$_{_{P-1}}$};

    \node [pythonred] at (4.125, 2.425) {$W_{1,1}$};
    \node [pythonred] at (4.275, 1.925) {$W_{2,1}$};
    \node [pythonred] at (4.250, 1.250) {$W_{P,1}$};
    \node [pythonblue] at (3.54, -0.56) {$W_{1,Q}$};
    \node [pythonblue] at (4.2, -1.000) {$W_{2,Q}$};
    \node [pythonblue] at (4.125, -1.69) {$W_{P,Q}$};
    \end{tikzpicture}
    \end{center}
        \caption[Diagram of a GPRN for RVs and activity indicators]{%
        Diagram of a GPRN applied to p datasets consisting of RVs and $p-1$
        activity indicators (AI). For example, AI$_1$ could be the FWHM and
        AI$_2$ the BIS. The regression network is formed by $Q$ nodes
        [$f_{1}(t)$, $f_{2}(t)$, ..., $f_{Q}(t)$] and $P\times Q$ weights
        [$W_{11}(t)$, ..., $W_{P1}(t)$, ..., $W_{1Q}(t)$, ..., $W_{PQ}(t)$]. For
        simplicity, we dropped their dependence on $t$ in the diagram, but each
        $f$ and $W$ follows an independent GP prior.}
        \label{fig:GPRN_RVs}
\end{figure}
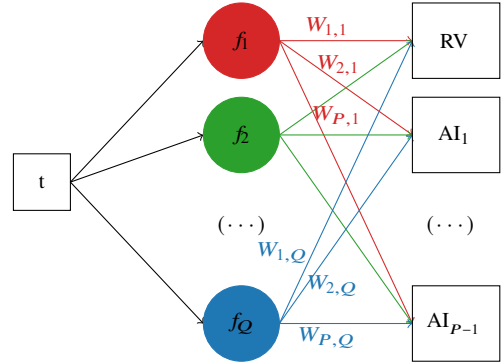

In our application, we use the GPRN to model RVs together with $P-1$ stellar
activity indicators (AI), with identical timestamps. This scheme is exemplified
in the diagram of Figure \ref{fig:GPRN_RVs}. The main objective of this
framework is to use the information contained in the activity indicators to help
characterise better the impact of activity in the RVs, thus enabling a cleaner
identification of any planetary signals that may be present. Again, each of the
nodes and weights in Figure \ref{fig:GPRN_RVs} are functions of time, $t$. This
allows for a fit to the data that can accommodate deviations from the stationary
expectation coming from a node. If the model incorporates several nodes, the
time-varying weights make it possible for each node to have a different impact
throughout the data. 

Henceforth, we will only consider the simplest possible configuration for a GPRN, with just one node plus one weight per time series. Such a model should be capable of capturing the quasi-periodic behaviour of the active regions on the surface of a star with the node, while the weights describe the variations associated with a longer-term magnetic cycle. This GPRN configuration is also the closest possible to the GP framework used in many previous works, thus it constitutes a natural starting point for the exploration of the GPRN capabilities.

\subsubsection{Variational inference}
\label{ch:GPRN_implementedFramework}

In contrast to a standard single-output GP, inference with a GPRN model is
intractable due to the lack of analytical expressions for the joint posterior
distribution with respect to the latent functions $\bm{f}(t)$ and $\bm{W}(t)$.
In order to overcome this, \citet{Wilson2012_GPRN} originally proposed two
approaches based on variational Bayes and MCMC. \citet{Nguyen2013b_GPRN} later
developed more efficient methods based on variational inference
\citep{Jordan1999_variationalInference}. Their aim was to find, among a family of
factorised distributions, the closest approximating distribution
$q(\bm{f},\bm{W})$ to the true posterior $p(\bm{f},\bm{W}|D)$. The closeness of
the approximation is measured through the Kullback-Leibler (KL,
\citealt{Kullback59}) divergence between $q(\bm{f},\bm{W})$ and
$p(\bm{f},\bm{W}|D)$, 
\begin{equation}
\label{eq:KLdivergence}
    \rm{KL} \left[q(\bm{f},\bm{W}) \, || \, p(\bm{f},\bm{W}|D) \right]
    = \mathbb{E}\left[ log\frac{q(\bm{f},\bm{W})}{p(\bm{f},\bm{W}|D)} \right],
\end{equation}
where $\mathbb{E}$ stands for expectation or expected value. One particular
family of factorised distributions is the so-called mean-field approximation,
which takes the form
\begin{equation}
\label{eq:Qs}
    q(\bm{f},\bm{W}) = \prod_{j=1}^{Q} q_f \left( f_j \right)  
                        \prod_{i=1}^{P} q_w \left(w_{ij} \right)
\end{equation}
\noindent 
where $q_f\left(f_j\right) = \mathcal{N}\left(\mu_{f_j}, \Sigma_{f_j} \right)$
and $q_w\left(w_{ij}\right) = \mathcal{N}\left(\mu_{w_{ij}}, \Sigma_{w_{ij}}
\right)$ are general Gaussian distributions. This factorisation allows for
analytical, alternating updates for the variational parameters $\mu$ and
$\Sigma$, whose expressions are provided in
Appendix~\ref{app:variationalParams}. 

Minimising the KL divergence in equation \eqref{eq:KLdivergence} is equivalent
to maximising the evidence lower bound (ELBO, \citealt{Blei2018_VarInference}),
which for the GPRN is given by
\begin{align}
    \label{eq:ELBO}
    \rm{ELBO}(q) =
        & ~ \mathbb{E}_q \left[ \log p( D | \bm{f}, \bm{W}) \right] 
          + \mathbb{E}_q \left[ \log p(\bm{f}, \bm{W}) \right] \nonumber  \\
        & + \mathcal{H}_q\left[ q(\bm{f}, \bm{W}) \right], 
\end{align}
\noindent
where the first term is the expected log-likelihood, the second is the expected
log-prior and the last is the entropy of $q(\bm{f}, \bm{W})$. For the full
Gaussian mean-field approximation, presented in expression \eqref{eq:Qs}, the
three terms in equation \eqref{eq:ELBO} can be computed in closed form. The
expressions were provided by \citet{Nguyen2013b_GPRN} and are reproduced in
Appendix~\ref{app:ELBO}. 

The expected values of the outputs $i$ for a new input time $t^\star$ are also
provided in~\cite{Nguyen2013b_GPRN}. This so-called predictive mean for the
outputs $\bm{y}_i^\star(t^\star)$, considering that the nodes and weights are independent, is given by
\begin{equation}
\label{eq:predictiveMean}
\begin{split}
\mathbb{E}\left[\bm{y}_i^\star\left( t^\star \right)\right]
    &~= \sum_{j=1}^{Q} \mathbb{E}\left[W_{ij}^\star \right] 
                       \mathbb{E}\left[f_{j}^\star \right] \\
    &~= \sum_{j=1}^{Q} K_{w_{ij}}^\star 
                       K_{w_{ij}}^{-1} \mu_{w_{ij}} K_{f_j}^\star K_{f_j}^{-1} \mu_{f_j}.
\end{split}
\end{equation}
\noindent
In this equation the predictive mean for each $y_i^\star(t^\star)$ is
obtained by summing the expected value of each node $f_{j}$ multiplied by the
expected value of the weight $W_{ij}$ that connects $f_{j}$ to $y_i$. 

The predictive variances for the outputs $i$ can be calculated through \citep{Wilson2012_GPRN}
\begin{equation}
\label{eq:predictiveCovariance}
\begin{split}
\mathbb{V}\left[ \bm{y}_i^\star \left( t^\star \right) \right]
= \sum_{j=1}^{Q}\Big[ &~ \mathbb{E}\left[W_{ij}^\star \right]^2 \mathbb{V}\left[f_j^\star\right] \\
&~+ \mathbb{V}\left[W_{ij}^\star \right] \mathbb{E}\left[f_j^\star \right]^2 \Big] + \sigma_{i}^{2},
\end{split}
\end{equation}
\noindent
where $\mathbb{V}$ stands for variance, with respect to either the node $f_j$ or
the weight $W_{ij}$, while $\sigma_{i}$ is the jitter term mentioned in
equation~\ref{eq:sigmaCalculation}. Note that this variance is with respect to
the predictive outputs of the GPRN model, not the predictive values for a
measurement made at $t^\star$. Although the predictions are the same, the
associated variance differs due to the extra uncertainty associated with the
measurement process. In particular, the expression for the variance associated
with the predictive values for the HARPS-N measurements made at the observed
times differs from what is provided by equation \ref{eq:predictiveCovariance}
insofar that instead of $\sigma_{i}$ one should use $\sigma_{y_{i}}$ as given by
equation \ref{eq:sigmaCalculation}.

We tested the implementation and accuracy of the mean-field approximation by comparing it with the MCMC approach proposed by \citet{Wilson2012_GPRN}, which relies on Elliptical Slice Sampling \citep{murray2010_ESS} to characterise the GPRN posterior. We found that both approaches provide fully compatible results for the predictive distributions. The MCMC algorithm is, however, much more computationally demanding.

The expressions for mean-field inference in the context of a GPRN, applied to
RVs and activity indicators, were implemented in a Python package called
\texttt{gpyrn}, that is made publicly
available\footnote{~\url{https://github.com/iastro-pt/gpyrn}}. The
implementation is computationally simple but could benefit from further
optimisation to speed up some linear algebra calculations. The run-time is
dominated by the inversion of $N\times N$ covariance matrices for the node and
weight GPs which, as currently implemented, scales with $N^3$. In principle,
some of these calculations could use more scalable approaches such as those
implemented in the \texttt{celerite} \citep{celerite} or \texttt{S+LEAF}
\citep{Delisle2022} packages, for example. The results presented in
section~\ref{ch:AnalysisSamples} below were obtained on a standard laptop
computer with an Intel\textsuperscript{\textregistered}
Core\texttrademark\xspace i7-6700HQ CPU, while for the more demanding analyses
of section \ref{ch:Sun} we used a computer cluster equipped with a 24-core
Intel\textsuperscript{\textregistered} Xeon\textsuperscript{\textregistered}
E5620 CPU.

\section{Application to simulated data}
\label{ch:AnalysisSamples}

\begin{table*}
    \centering
    \begin{tabular}{lll c c c c}
    \hline\\[-2ex]
    Hyper-parameter & Description & Prior & %
    Input & Units \\
    \hline\\[-1ex] 
    $\eta^n_2$ &  Decaying timescale & %
        $\mathcal{L\,U} \left(\tav, 10\,\tspan \right)$ %
        & 17&  days \\[1ex]
    $\eta^n_3$ & Period & %
        $\mathcal{U} \left( 10, 50 \right)$ %
        &23 &   days \\[1ex] 
    $\eta^n_4$ & Lengthscale & %
        $\mathcal{LU} \left( 0.1, 5 \right)$ %
        &0.75  \\[1ex] 
    \multirow{2}{*}{$\eta^w_1$} & 
    \multirow{2}{*}{Amplitude} & 
    \multirow{2}{*}{$\mathcal{MLU} \left( y_{\sigma}, 2\,y_{\rm{ptp}} \right)$} &
          7 &  m/s \\[1ex] 
    & & & 7 &  m/s \\[1ex] 
    \multirow{2}{*}{$\eta^w_2$} & 
    \multirow{2}{*}{Decaying timescale} & 
    \multirow{2}{*}{$\mathcal{LU} \left(\tav, 10\:\tspan \right)$} %
            & 29  & days \\[1ex]
        & & & 109 &  days \\[1ex]
    \multirow{2}{*}{s} & 
    \multirow{2}{*}{White noise amplitude} & 
    \multirow{2}{*}{$\mathcal{MLU} \left( y_{\sigma}, 2\,y_{\rm{ptp}} \right)$} & 
        0 &  m/s \\[1ex] 
    & & & 0& m/s \\[1ex]
    \hline\\
    \end{tabular}
    \caption{Hyper-parameter input values for the simulations, and assumed 
    priors for the GPRN analysis of the simulated data. The superscripts on some 
    hyper-parameters identify if they belong to the node ($n$) or to the weights ($w$). 
    In the case of the prior distributions, $\mathcal{U}(a,b)$ denotes a uniform
    distribution between $a$ and $b$, $\mathcal{LU}(a,b)$ a log-uniform
    distribution between $a$ and $b$, while $\mathcal{MLU}$ is a modified
    log-uniform distribution with \textit{knee} $a$ and upper limit $b$
    \textbf{\citep[see eqn. 16 in][]{Gregory2005}}. Also, \tav is the average time between
    consecutive observations and \tspan is the timespan of the observations,
    while $y_\sigma$ and $y_{\rm{ptp}}$ are the standard deviation and
    peak-to-peak difference for a given vector of observations $y$,
    respectively. We fix the amplitude of the node, $\eta^n_1$, to 1 m/s.}
    \label{table:01_Priors}
\end{table*}

In order to test the GPRN implementation, we simulated twenty sets of
50 observations each using a GPRN with one node and two different weights. The
node was assumed to have a quasi-periodic covariance function, as given by
Equation \ref{eq:QPkernel}, with an amplitude $\eta^n_1=1$ m/s, a decaying
timescale $\eta^n_2=17$ days, a period $\eta^n_3=23$ days, and $\eta^n_4=0.75$.
Each weight was assumed to have a squared exponential covariance function, 
\begin{equation}
\label{eq:SEkernel}
    k_{\mathcal{SE}}\left(t,t'\right) = 
        \eta^{2}_{1} \exp \left[ 
            - \frac{\left(t-t'\right)^2}{2\eta^{2}_{2}} \right],
\end{equation}
with the same variance $\eta^w_1=7$ m/s, but different decaying timescales
$\eta^w_2$ of $29$ days and $109$ days. Both node and weights had mean functions
equal to zero. Henceforth, the node and weight hyper-parameters are identified
by the superscripts $n$ and $w$, respectively.

The times for the observations were randomly drawn from an uniform distribution
between 0 and 150 days. These input times are shared by both simulated datasets,
whose outputs were obtained by multiplying one sample from the GP prior for the
node with one independent sample from the GP prior for each weight. In order to
make these outputs resemble more closely real observations, we added to each a
simulated measurement error drawn from a normal distribution with mean equal to
0 m/s and standard deviation equal to 1 m/s. Note that the GPRN model used in the
simulation does not contain any white noise process, i.e. we set the jitter
terms to zero, namely in equation~\ref{eq:sigmaCalculation}.

We then analysed the simulated observations, expecting to recover, on
average, the assumed values for the GPRN hyper-parameters as well as the latent
functions. The GPRN model used in the analysis was the same as for the
simulations, except that the weight variances $\eta^w_1$ were assumed different
and we included a free jitter hyper-parameter per dataset to account for
possible Gaussian white noise. The uncertainties associated with the
simulated measurements were assumed known a priori and added in quadrature to
those jitter terms, as described through equation~\ref{eq:sigmaCalculation}. In
order to simplify the analysis, we fixed the mean functions to zero, and the
amplitude of the node, $\eta^n_1$, to its fiducial value of 1 m/s. The prior
distributions are listed in Table \ref{table:01_Priors} and we use
\texttt{emcee} \citep{ForemanMackey2019_emcee} to characterise the posterior
distribution for the model hyper-parameters. We set the number of MCMC walkers
to twice the number of hyper-parameters and run the chains until a convergence
criteria of 25 times the integrated autocorrelation time, $\tau$, is achieved.
This criterion defines $\tau$ as the number of steps needed for a chain to
forget where it began or reach equilibrium \citep{Sokal1997_ACT}. We calculated
the value of $\tau$ every 5000 iterations. We then discarded the $2\tau$ initial
samples as burn-in.


\begin{figure*}
    \centering
    \includegraphics[width=0.8\textwidth]{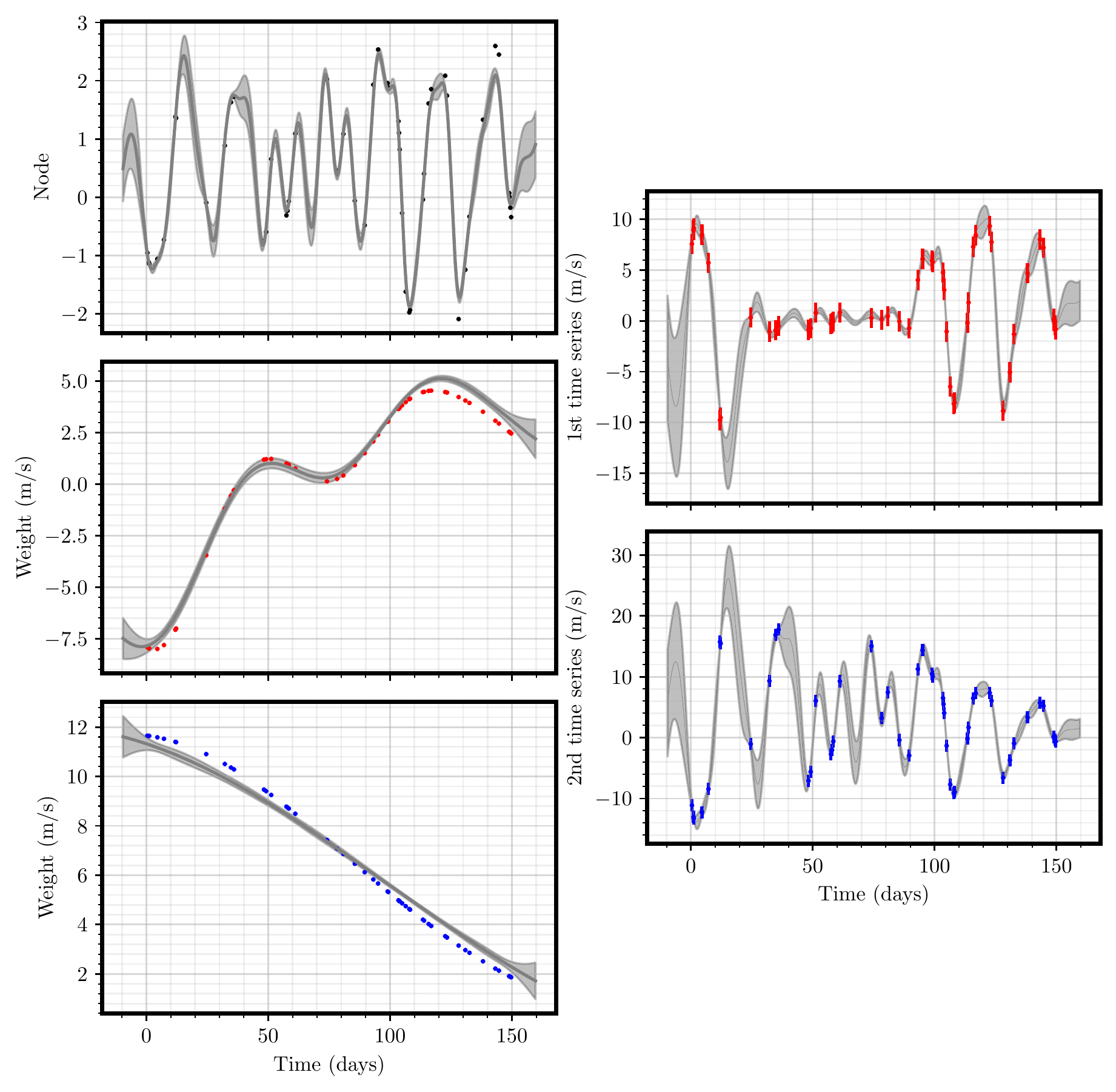}
    \caption{The left panels show the simulated values (dots)
    and the posterior predictive means (black lines) for the node and 
    the weights connected to the first and second time series, assuming 
    the inferred MAP values for the GPRN hyper-parameters. The grey band 
    around each line represents the square root of the posterior predictive 
    covariance. The right panels show the same quantities with respect to the 
    simulated outputs, as well as the uncertainties associated with the later.}
    \label{fig:samplesTimeseries_fullmodel}
\end{figure*}

Figure \ref{fig:samplesTimeseries_fullmodel} displays the results for
one of the simulated datasets, showing the predictive means for the outputs, the
node, and the weights assuming the MAP values for the GPRN hyper-parameters. As
expected the GPRN manages to capture the structure of both time series and also
of the latent node and weights. While for individual datasets the recovered
hyper-parameters can deviate from the input values, the average results from the
twenty simulations show no significant biases, except for the decaying
timescales of both weights which tend to be overestimated. This suggests the
node was capable of modelling some of the long-term variations present in the
simulated data, leading to a relative increase in the posterior probability of 
large values for the timescales of the weights.

\section{Sun as a star observations}
\label{ch:Sun}

\begin{figure*}
    \begin{center}
        \includegraphics[width=0.9\textwidth, keepaspectratio=true]{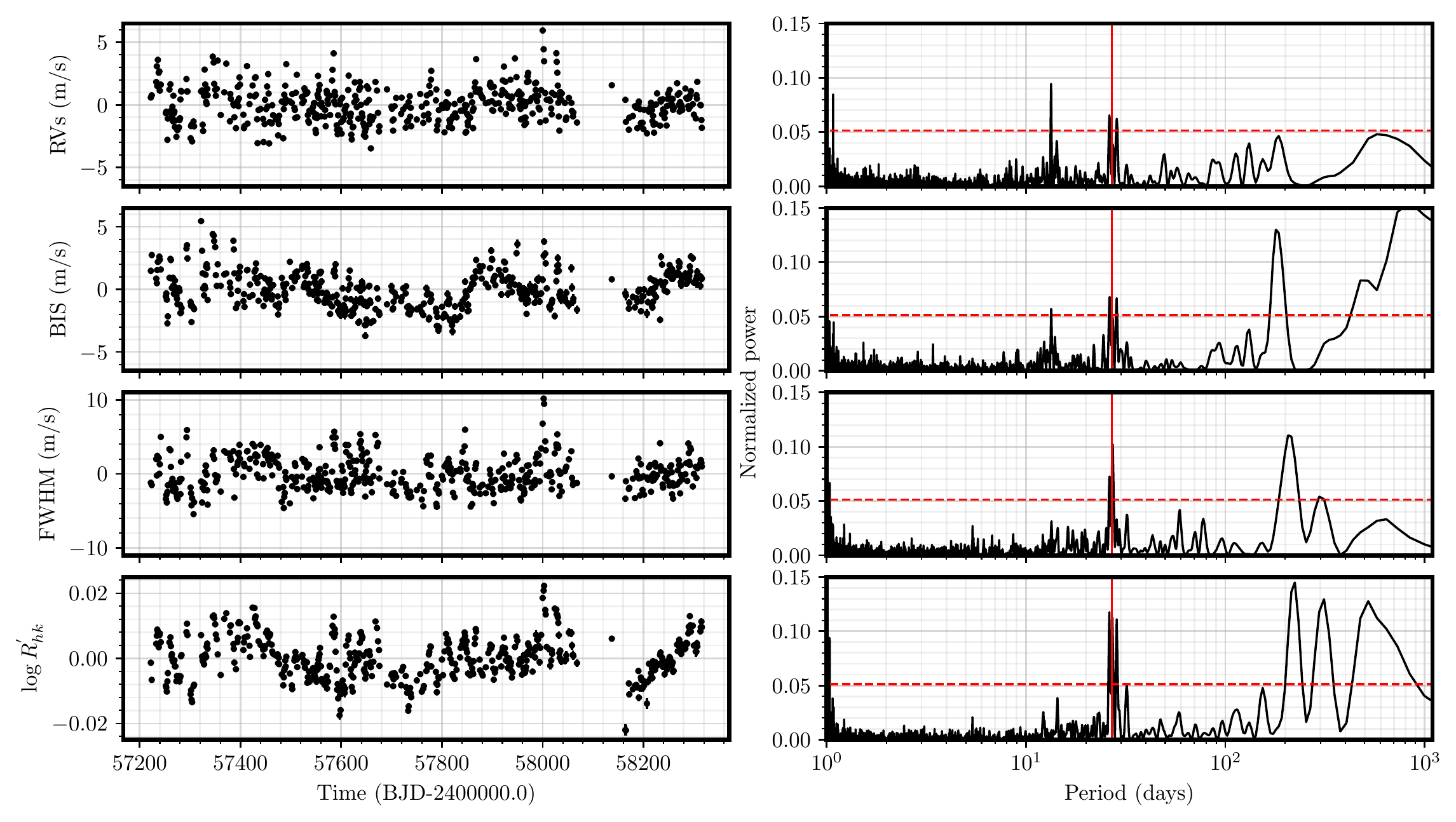}
    \end{center}
    \caption{The HARPS-N solar telescope RV, BIS, FWHM, and \logRhk 15-minute
     averaged measurements we consider in our analysis are displayed on the left
     after subtracting the best-fit linear trend from the data. The GLS
     periodograms with respect to these measurements are shown on the right. The
     vertical solid lines mark the 27-days period, while the horizontal dashed
     lines show the false alarm probability (FAP) of $1\%$.}
    \label{fig:sunMeasurements}
\end{figure*}

The Sun has been continuously studied and monitored over millennia, enabling us
to determine its characteristics with high precision. It is a typical
main-sequence star of spectral type G2V with a radiative interior and a
convective shell \citep{DelZanna2013_sunInterior}. The solar dynamo is the
physical process  responsible for the magnetic activity that gives rise to the
stellar signals that appear in RV measurements
\citep[e.g.][]{Pagano2013_stellarActivity, Charbonneau2014_solarDynamo}. The
level of magnetic activity changes with time, leading to a 11-year sunspot
cycle, during which the number of spots, flares and coronal mass ejections
varies \citep[see e.g.][]{Hathaway2015_solarCycle}. 

In this work, we analyse three years of Sun-as-a-star observations from the
HARPS-N solar telescope~\citep{Dumusque2015_HARPS-N}. The observations were
obtained between 2015 and 2018, during the declining phase of solar cycle 24.
The data was first presented and analysed by \citet{Cameron2019_SunAsAStar}.
More recently, \citet{Dumusque2021_SunRVs} re-visited the data after adapting
the ESPRESSO DRS \citep{Pepe2021_ESPRESSO_DRS} to the HARPS-N spectrograph, and
re-estimated the solar RVs. They adopted a more stringent criteria to evaluate
the quality of the individual spectra, obtaining a decrease in the day-to-day RV
scatter and a stronger correlation between the RVs and activity
indicators. The full set of spectra, CCFs, and RV time series are publicly
available through the DACE platform\footnote{\url{https://dace.unige.ch/sun}}.

In total, there are 34550 RV measurements available. To simulate a standard
15-minute exposure for solar-type stars, which helps remove the RV contributions
from oscillations and granulation, we randomly selected three consecutive
measurements per day (whenever possible) and averaged them, discarding all other
data points. The RV and \logRhk uncertainties were determined by adding
quadratically the uncertainties for each set of three measurements, i.e. they
were calculated by taking the square root of the average of these uncertainties
squared and then dividing by three. The uncertainties associated with the BIS
and FWHM averaged measurements were then defined as twice the value of the RV
uncertainties. This leads to a total of 497 RV, FWHM, BIS, and \logRhk averaged
measurements, spanning 1094 days (see Figure \ref{fig:sunMeasurements}). Since
the effects of the Solar System planets were already removed from the RVs, these
time series can be assumed to contain only the imprint of solar activity, on
timescales exceeding 15 minutes, and are thus an excellent test set for applying
the GPRN.

\subsection{Periodogram analysis}
\label{ch:PeriodogramAnalysis}

We would like to check for the presence of periodic signals in the solar data
before using the GPRN framework to analyse them. Thus, we computed the
generalised Lomb-Scargle (GLS, \citealt{Zechmeister2009, Astropy2018})
periodogram for each of the assembled RV, BIS, FWHM, and \logRhk time series,
after the best-fit linear trend is removed. These can be seen on the right
panels of Figure \ref{fig:sunMeasurements}.

All periodograms show a forest of peaks around 27 days, the synodic rotation
period of the Sun \citep{Wilcox1972_SunSynodicPeriod}. These peaks are
significant (i.e. higher than the 1\% false alarm probability) in all four time
series. However, in the RVs the most significant peak is at the first harmonic
of the rotation period, around 13.5 days. The BIS shows the periodogram
structure closest to that of the RVs, also with a significant peak around 13.5
days but with lower power than the 27 days peak. On the contrary, the 13.5 days
peak does not appear as significant in the FWHM and \logRhk periodograms. At
longer periods, all time series show periodogram peaks around 200 days, most
noticeable in the activity indicators than in the RVs. This is probably related
to secular trends due to smooth long-term changes in the CCF area
\citep{Cameron2019_SunAsAStar}.

The periodograms clearly show that both the RVs and the activity indicators
contain signals induced by activity, which have different structure in each
time series. We note that this periodogram analysis serves just as a data
exploratory step, and will only loosely inform our prior assumptions when
modelling the data.

\subsection{Setup for the GPRN analysis}
\label{ch:Analysis}

\begin{table*}
    \centering
    \begin{tabular}{lll}
    \hline\\[-2ex] 
    Parameter & Description (units) & Prior \\
    \hline\\[-1ex] 
    $\eta^n_1$ &  Amplitude (m/s) & $\mathcal{MLU} \left( \left[\text{y}_{RV}, \text{y}_{AI} \right]_{\sigma}^{\text{min}}, 2\left[ \text{y}_{RV}, \text{y}_{AI}\right]_{\text{ptp}}^{max} \right)$ \\[1ex] 
    $\eta^n_2$ &  Decaying timespan (days) 
        & $\mathcal{L\,U} \left(\tav, 10\,\tspan \right)$ \\[1ex] 
    $\eta^n_3$ &  Period (days) 
        & $\mathcal{U} \left( 10, 50 \right)$ \\[1ex] 
    $\eta^n_4$ & Length scale 
        & $\mathcal{L\,U} \left(0.1, 5 \right)$ \\[2ex] 
    $\eta^w_1$ & Amplitude (m/s)
        & $\mathcal{M\,L\,U} \left( y_{\sigma}, 2\,y_{\rm{ptp}} \right)$ \\[1ex]
    $\eta^w_2$ & Decaying timescale (days) & 
        $\mathcal{L\,U} \left(\tav, 10\,\tspan \right)$ \\[2ex] 
    s & White noise amplitude (m/s) 
        & $\mathcal{M\,L\,U} \left( y_{\sigma}, 2\,y_{\rm{ptp}} \right)$ \\[2ex] 
    slope & Slope of the mean function (m/s/day) & 
        $\mathcal{N} \left( 0, y_{\sigma} / t_{\sigma} \right)$\\[1ex] 
    offset & Offset of the mean function (m/s) 
        & $\mathcal{U} \left( y_{\rm min}, y_{\rm max} \right)$\\[1ex] 
    \hline\\
    \end{tabular}
    \caption{GPRN hyper-parameters and associated prior distributions used in
    the analysis of the HARPS-N solar observations. The $y_{RV}$ and $y_{AI}$
    represent, respectively, the observed RVs and activity indicators (FWHM,
    BIS, or \logRhk), while $t$ denotes the times of the observations. We denote
    by \tav the average time between consecutive observations, \tspan represents
    the timespan of the observations, while $y_\sigma$ and $y_{\rm{ptp}}$ are
    the standard deviation and peak-to-peak difference for a given vector of
    observations $y$, respectively. $\mathcal{U}(a,b)$ stands for a uniform
    distribution between $a$ and $b$. $\mathcal{LU}(a,b)$ represents a
    log-uniform distribution with shape parameters $a$ and $b$. $\mathcal{MLU}$
    denotes the modified log-uniform distribution, a log-uniform distribution
    whose support includes zero. Note that for the GPRN joint analysis of the
    RVs and an activity proxy, the $\eta^n_1$ prior was defined using the
    smallest $y_\sigma$ and $y_{\rm{ptp}}$ among the two datasets.}
    \label{table:01_Priors2datasets}
\end{table*}

We proceed with the application of the GPRN framework to the solar observations,
first modelling the individual time series and then jointly pairing  the RVs
with each activity indicator. Our aim is to determine whether the GPRN is
capable of modelling the RVs jointly with the activity indicators as well as
when they are modelled individually. The closer the results, the higher will be
the information carried by the given activity indicator about the impact stellar
activity has on RVs. 

As already mentioned, we will always consider the simplest possible GPRN model, 
with just one node and thus one weight per dataset analysed, each having 
associated hyper-parameters. The node is assumed to have a quasi-periodic covariance
function, while each weight has a squared exponential covariance function with
specific hyper-parameters. As before, the node and weight hyper-parameters are
identified by the superscripts $n$ and $w$, respectively.

The prior distributions used in all analysis are shown in Table
\ref{table:01_Priors2datasets}. Whenever two datasets were jointly analysed the
prior for the node amplitude, $\eta_1$, was adapted: the knee of the modified
log-uniform prior (the point where the distribution changes from uniform to
log-uniform) is now defined as the lowest standard deviation of the two time
series, denoted by $ \left[\text{y}_{RV},
\text{y}_{AI}\right]_{\sigma}^{\text{min}}$. The upper bound of this prior
distribution is specified to be twice the largest of the two peak-to-peak
amplitudes obtained from the RVs and the activity indicator being considered,
and represented by $\left[ \text{y}_{RV},\text{y}_{AI}\right]_{\text{ptp}}^{max}$. 
The prior for the $\eta_1$ associated with each weight only uses the standard 
deviation and peak-to-peak amplitude of the dataset the weight is related to.

The $\eta_2$ parameters for the node and weights were assigned broad log-uniform
priors between the mean time difference between consecutive points, \tav, and
ten times the full timespan of the data, \tspan. In addition, to recover more
physically-meaningful values \citep[see, e.g.,][]{Kosiarek2020_photometryAsProxy}, 
the value for the $\eta_2$ associated with the node, $\eta^n_2$, 
was a priori constrained to have values higher than
half the value of $\eta_3$. This also ensures $\eta_3$ can be interpreted as a
period. The weights $\eta_2$ values, $\eta^w_2$, are also assumed a priori to be
higher than $\eta^n_2$. These conditions ($0.5\,\eta_3 < \eta^n_2 < \eta^w_2$)
should increase the chances the node will model the rotation-induced signals created 
by active regions, while the weights model their evolution on the longer timescales related
to the solar magnetic cycle. Favouring a priori slowly varying weight functions 
will also reduce the risk of overfitting the data.

A linear mean function was associated to each time series, parameterised by a
slope and an offset. For the slope, we assign a Gaussian prior with zero mean
and standard deviation equal to the ratio of the standard deviations of the
measurements and the timestamps. A uniform prior between the minimum and maximum
values of the respective dataset was used for the offset. Finally, each dataset
is also associated with a jitter term, $s$, which models additional stationary
Gaussian white noise not accounted for by the observational uncertainties. Each
jitter is assigned a modified log-uniform prior.

Similarly to the analysis of Section~\ref{ch:AnalysisSamples}, we use
\texttt{emcee} to explore the parameter space. The same setup is used, with the
number of walkers equal to twice the number of hyper-parameters, and the same
convergence criteria applied as before. 

\subsection{Results}

We will first describe the results of the analysis of each dataset individually.
The GPRN model used here, with just one node and one weight, approximates a GP
in the limit of a constant weight (or a very large $\eta_2^w$). The difference
then resides in the shape of the joint posterior with respect to the outputs,
which in the GP case is a multivariate Gaussian, while in the GPRN case is
generally a multivariate distribution with longer tails. The weight in the GPRN
provides some additional flexibility to model non-stationary changes in the
solar activity signals over the three years of observations. The results of an
analysis with a GPRN with one node and one weight can thus be compared to some
extent with those obtained using a GP model \citep{Langellier2020_detectionLimitsGPs}.

Afterwards, we present the results of the joint analysis of RVs and each of the
activity indicators, again assuming only one node in the GPRN. In this case, the
node will describe the assumed shared quasi-periodic activity-related latent
process.

For each analysis, we quote the \textit{maximum a posteriori} (MAP) value for
the joint posterior distribution of the hyper-parameters, as well as the medians
and $68\%$ credible intervals (i.e. the $16\%$ and $84\%$ quantiles) with
respect to all marginal posterior distributions. These values are shown in
Tables \ref{table:Results1} and \ref{table:Results2}. The corner plots showing
the marginal, single and pairwise, posterior distributions for all
hyper-parameters are in Appendix \ref{appendix:corners_sun}. Most of these distributions are well defined and localised. The posterior distributions for the $\eta_1$ parameters are exceptions, due to the strong anti-correlation between the node and weight values for those parameters, whose product must equal the variance per dataset to a first approximation. Further, the posterior distributions for some weight GP timescales ($\eta^w_2$) are ill-defined for very large values of these parameters, in the sense that they just follow the behaviour of the prior distribution for such values. This is because the data does not contain information that can be used to distinguish between them.

\begin{table*}
    \centering
    \begin{tabular}{llcccccccc}
    \hline\\[-2ex]
    &
    & \multicolumn{2}{c}{RV} 
    & \multicolumn{2}{c}{BIS} 
    & \multicolumn{2}{c}{FWHM} 
    & \multicolumn{2}{c}{\logRhk} \\[1ex]
    \hline\\[-2ex] 
    \multicolumn{2}{l}{Parameter} & MAP & Median & MAP & Median & MAP & Median & MAP & Median\\
    \hline\\[-1ex] 
    $\eta_1^n$ &        & 0.720     & $1 ^{+5} _{-1}$           & 1.568     & $2 ^{+5} _{-1}$            & 0.647     & $2 ^{+6} _{-1}$           & 0.010      & $0.05 ^{+0.04} _{-0.02}$            \\[1ex]
    $\eta_2^n$ & [days] & 19.972    & $20.3 ^{+2.0} _{-1.9}$          & 23.235    & $23.6 ^{+1.6} _{-1.6}$           & 19.001    & $19.1 ^{+1.6} _{-1.8}$          & 20.530     & $20.5 ^{+1.4} _{-1.4}$           \\[1ex]
    $\eta_3^n$ & [days] & 25.921    & $26.2 ^{+0.7} _{-0.6}$          & 26.853    & $27.0 ^{+0.4} _{-0.4}$           & 28.244    & $28.0 ^{+0.6} _{-0.6}$          & 28.273     & $27.4 ^{+0.7} _{-0.7}$           \\[1ex]
    $\eta_4^n$ &        & 0.584     & $0.61 ^{+0.07} _{-0.06}$           & 0.724     & $0.73 ^{+0.06} _{-0.05}$            & 0.731     & $0.72 ^{+0.09} _{-0.09}$           & 1.074      & $1.05 ^{+0.09} _{-0.08}$            \\[1ex]
    $\eta_1^w$ & [m/s]  & 1.038     & $2 ^{+5} _{-1}$           & 0.855     & $1 ^{+5} _{-1}$            & 1.898     & $2 ^{+6} _{-1}$           & 0.102      & $0.05 ^{+0.04} _{-0.03}$            \\[1ex] 
    $\eta_2^w$ & [days] & 1001.229  & $3569 ^{+4026} _{-2123}$  & 2431.229  & $4909 ^{+3589} _{-2517}$   & 1812.053  & $4662 ^{+3829} _{-2618}$  & 4124.169   & $5519 ^{+3541} _{-2991}$   \\[1ex]
    $\text{s}$ & [m/s]  & 0.781     & $0.79 ^{+0.04} _{-0.04}$           & 0.414     & $0.418 ^{+0.030} _{-0.030}$            & 1.136     & $1.13 ^{+0.06} _{-0.06}$           & 0.001      & $0.00100 ^{+0.00009} _{-0.00008}$              \\[1ex]
    \hline\\
    \end{tabular}
        \caption{Results for the GPRN analysis of the datasets individually. The
         superscripts on some hyper-parameters identify if they belong to the
         node ($n$) or to the weights ($w$): $\eta_1^n$, $\eta_2^n$ and 
         $\eta_3^n$ control, respectively, the amplitude, the decay timescale and 
         the periodicity of the correlations between the node outputs, while $\eta_4$ 
         defines their timescale of variation relative to $\eta_3$; $\eta_1^w$ and 
         $\eta_2^w$ control, respectively, the amplitude and the decay timescale of 
         the correlations between each weight outputs. The first column identifies each
         hyper-parameter, while each subsequent pair of columns show the MAP
         values associated with the joint posterior distribution of the GPRN
         hyper-parameters and the median values plus 16th and 84th percentiles
         (68\% credible intervals) with respect to the marginal posterior
         distributions of each GPRN hyper-parameter, in the case of the RVs,
         BIS, FWHM and \logRhk, respectively.}
        \label{table:Results1}
\end{table*}

\subsubsection{Individual analysis}

The analyses of the four datasets individually, using the GPRN model previously
described, all lead to similar results for the values of the GPRN
hyper-parameters. In particular, we always recover with high precision the
synodic rotation period of the Sun, about 27 days, through $\eta^n_3$. All the
inferred timescales for the decaying rotation-induced correlations, $\eta^n_2$,
are close, within 19 to 23 days, and consistent taking into account their
associated $68\%$ credible intervals. The very high inferred values for the
timescales of the weight functions, $\eta^w_2$, suggest that the time series
behave in a quasi-stationary manner. This can be clearly seen through the slowly
varying predictive means for the weights, shown on
appendices~\ref{fig:RV_fullPlots},~\ref{fig:BIS_fullPlots},~\ref{fig:FW_fullPlots},
and~\ref{fig:RHK_fullPlots}. Nevertheless, the fact that the weight functions
clearly show a decreasing trend, whichever the dataset considered, is an
indication of non-stationary behaviour due to diminishing data variance, as
expected of a star approaching its activity minimum.

\begin{figure*}
    \includegraphics[width=0.9\textwidth,height=0.9\textheight]{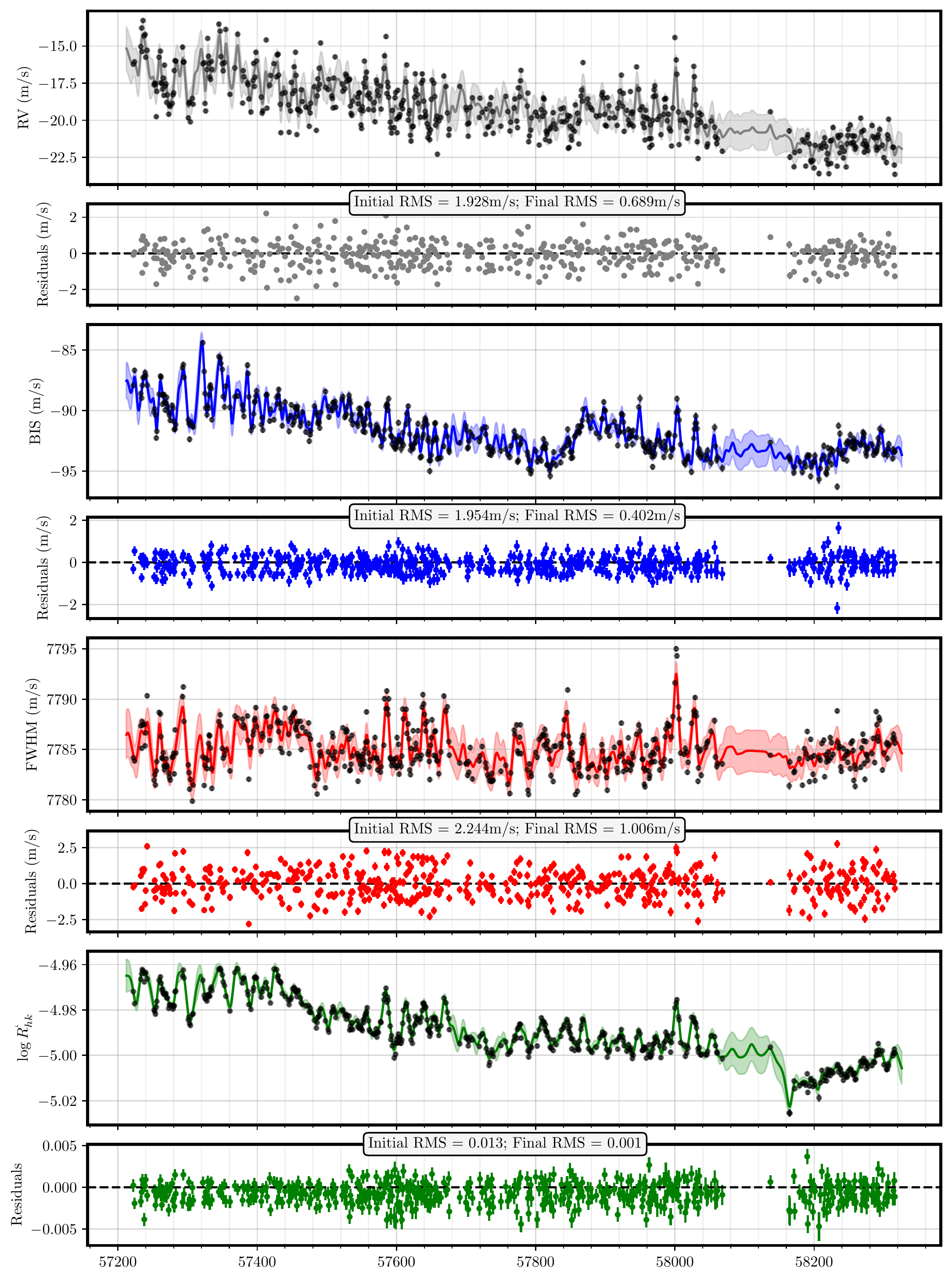}
    \caption{Visual representations of the predictive means and standard
    deviations associated with the MAP values for the GPRN hyper-parameters
    obtained when each considered time series is analysed individually. The
    panels show, from top to bottom, the results for the analysis of the RVs (in
    grey), the BIS (in blue), the FWHM (in red), and the \logRhk (in green), as
    well as the measured values and associated uncertainties (which are not
    always visible). We also show the residuals after subtracting the predictive
    means associated with each dataset, and quote the rms associated with the
    data shown (initial) and with the residuals (final).}
    \label{fig:individualAnalysis}
\end{figure*}

All four GPRN models managed to achieve very good fits to the data, as can be
seen in Figure \ref{fig:individualAnalysis}. Subtracting the MAP predictive mean
from each respective dataset leads to a residual rms significantly lower than
that of the original dataset. The RV rms was reduced by a factor of 2.8, from
1.928 m/s to 0.689 m/s. The decrease in the case of the FWHM was similar (by a
factor of 2.2), while for the BIS and \logRhk we find even higher rms reduction
factors of 4.9 and 9.7, respectively. These could indicate some degree of 
over-fitting, but all the residual rms values are higher than the average 
measurement uncertainties associated with each dataset, by factors of 5.2 (RVs), 
1.5 (BIS), 3.8 (FWHM) and 1.2 (\logRhk). This in fact suggests measurement 
uncertainties may have been underestimated, systematic effects are present 
(e.g. due to instrument miscalibration or atmospheric phenomena) and/or 
there are physical processes occurring in the Sun which are not being modelled 
adequately by the GPRN,  like granulation and super-granulation. The latter 
change on timescales of a few hours to days \citep{Meunier2019_SuperGranulation}, 
significantly smaller than the timescales associated with solar rotation and 
longer-term magnetic cycles.

\begin{figure}
    \includegraphics[width=0.48\textwidth]{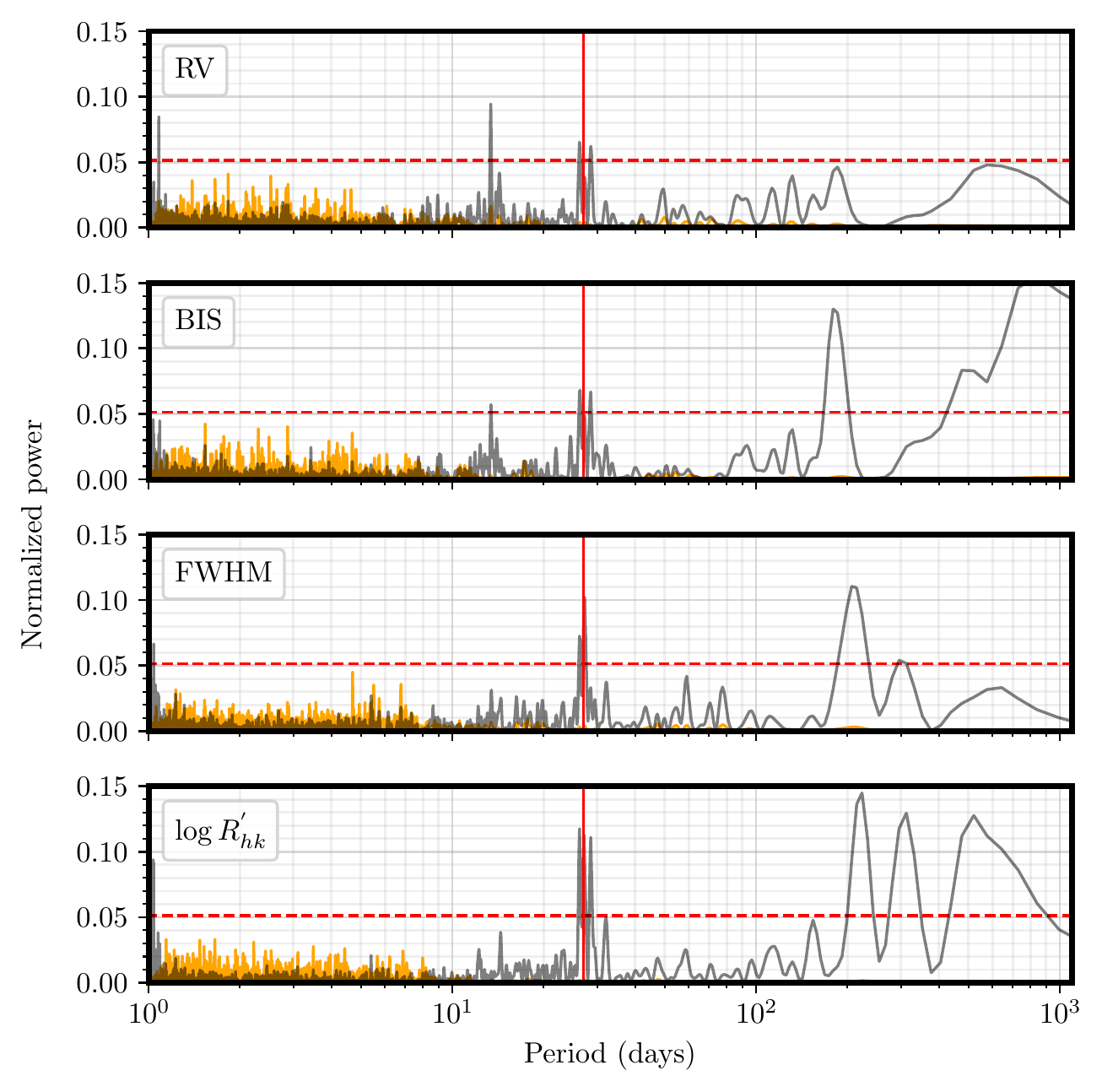}
     \caption{Comparison of the GLS periodograms of the time series depicted in
     Figure~\ref{fig:sunMeasurements} (in grey) and of their residuals after
     subtracting the MAP predictive means derived when each time series is
     analysed individually with the GPRN model considered (in orange). The
     vertical solid lines mark the 27-days period, while the horizontal dashed
     lines show the false alarm probability (FAP) of $1\%$ (all in red).}
    \label{fig:residuals_individual}
\end{figure}

In Figure \ref{fig:residuals_individual} we show the GLS periodograms for the
residuals after subtracting from each time series the respective MAP predictive
means. The periodograms of the original data (i.e. same as in Figure
\ref{fig:sunMeasurements}) are also shown for comparison. Clearly, the GPRN
model can describe essentially all the power present in the original time
series, with no significant peaks left in the periodograms of the residuals. The
model de-correlates the data to the point where the residuals resemble white
noise.

\subsubsection{RV and BIS}
\begin{figure*}
    \includegraphics[width=0.9\textwidth]{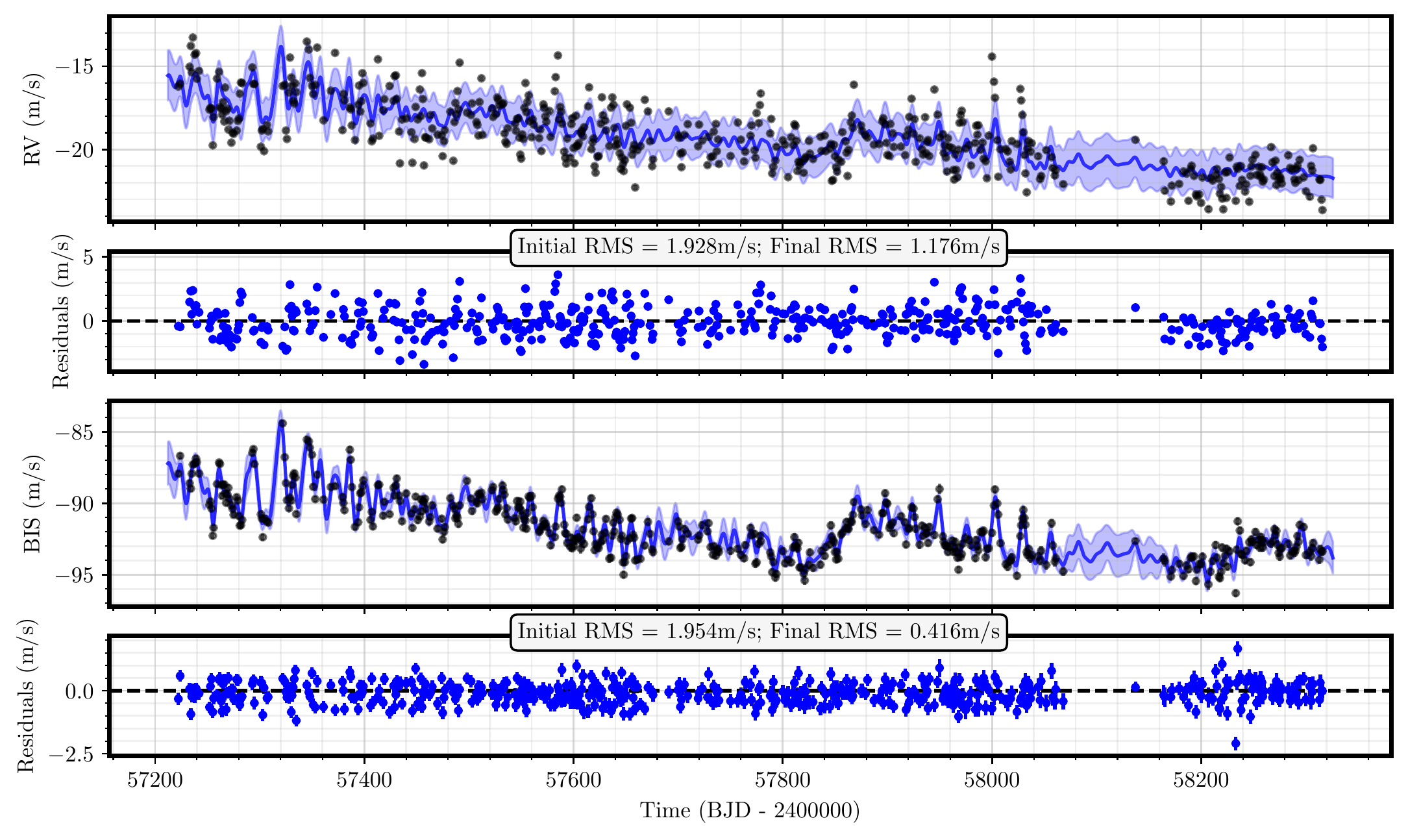}
    \caption{Predictive means and standard deviations associated with the MAP
    values for the GPRN hyper-parameters obtained when the RVs and BIS are
    analysed jointly. Measured values, residuals (after subtracting the
    respective predictive means) and associated uncertainties (which are not
    always visible) are also shown, as well as the rms associated with the
    (initial) data and with the (final) residuals.}
    \label{fig:RVsBISfit_GPRN}
\end{figure*}

The first joint analysis combined the RV and BIS measurements. Figure
\ref{fig:RVsBISfit_GPRN} shows the GPRN posterior predictive means and standard
deviations using the MAP values of the hyper-parameters. As expected, the
reduction in rms, from the original datasets to the sets of residuals obtained
by subtracting the MAP predictive means from the former, is now somewhat lower
than the reduction in rms obtained through the analysis of the datasets
individually. The RV and BIS rms are now reduced by a factor of 1.6 and 4.7,
instead of 2.8 and 4.9, respectively. This results from our imposition of a
common latent process behind the generation of the two datasets, whose
characteristics are described through the single node. Surprisingly, the node
seems to give preference to fitting the BIS over the RV measurements. This can
be seen in the significantly smaller degradation in the rms reduction, but also
in the much lower increase in the MAP value for the jitter, 2\% for the BIS
versus 52\% for the RVs, and in the posterior predictive uncertainties, with
respect to what was found in the individual analysis. This could be the result
of the somewhat lower harmonic complexity present in the BIS dataset, as can be
seen by comparing the BIS and RV periodograms.

The mode for the posterior marginal distribution for $\eta_3^n$ is very close to
27 days, the synodic rotation period of the Sun, but there are local maxima near
36-37 days and 44-45 days (see Figure \ref{fig:RVandBIS_gprn_corner}). These are
associated with increasingly smaller values for $\eta_4^n$, but do not seem
correlated with specific values of the other hyper-parameters. The estimate for
$\eta_2^n$ is consistent with the average lifetime of active regions being of a
few weeks, close to the solar rotation
period~\citep[e.g.][]{Hathaway2008_sunspotDecay,Driel2015_EvolutionOfActiveRegions}.

As was the case in the individual analysis of the datasets, the posterior
predicted means for the weight functions show a decrease during the timespan of
the observations. This results from the evolution of the solar magnetic cycle,
which was approaching an activity minimum. The clearly more structured weight
function in the case of the RVs (see Figure \ref{fig:fullPlots_RVBIS}) just
reflects the need to compensate for the degradation in the quality of the fit
provided by the node.

\begin{table*}
    \centering
    \begin{tabular}{lccccccc}
    \hline\\[-2ex] 
    &
    & \multicolumn{2}{c}{RV and BIS} 
    & \multicolumn{2}{c}{RV and FWHM} 
    & \multicolumn{2}{c}{RV and \logRhk} \\[1ex]
    \hline\\[-2ex] 
    Parameter && MAP & Median & MAP & Median & MAP & Median\\
    \hline\\[-1ex] 
    $\eta_1^n$    & &  1.163     & $0.5 ^{+1.4} _{-0.4}$            & 1.951     & $0.51 ^{+1.15} _{-0.34}$            & 0.001      & $0.07 ^{+0.05} _{-0.03}$            \\[1ex] 
    $\eta_2^n$    & [days] &  24.319    & $23.6 ^{+1.6} _{-1.6}$           & 22.183    & $20.8 ^{+1.9} _{-2.0}$           & 19.654     & $20.5 ^{+1.4} _{-1.4}$           \\[1ex] 
    $\eta_3^n$    & [days] &  26.745    & $27.1 ^{+0.4} _{-0.4}$           & 27.643    & $27.7 ^{+0.6} _{-0.6}$           & 28.615     & $28.3 ^{+0.7} _{-0.7}$           \\[1ex] 
    $\eta_4^n$    &        &  0.783     & $0.74 ^{+0.06} _{-0.06}$            & 1.008     & $0.86 ^{+0.12} _{-0.11}$            & 1.057      & $1.08 ^{+0.09} _{-0.08}$            \\[1ex] 
    RV $\eta_1^w$ & [m/s]  &  0.413     & $2 ^{+5} _{-2}$            & 0.440     & $2 ^{+4} _{-1}$            & 5.939      & $10 ^{+6} _{-4}$           \\[1ex]  
    RV $\eta_2^w$ & [days] &  281.890   & $2058 ^{+4117} _{-1677}$   & 37.319    & $66 ^{+56} _{-24}$         & 86.682     & $89 ^{+22} _{-19}$         \\[1ex] 
    AI $\eta_1^w$ & [m/s]  &  0.669     & $3 ^{+7} _{-2}$            & 0.675     & $5 ^{+9} _{-3}$            & 0.024      & $0.06 ^{+0.04} _{-0.03}$            \\[1ex]  
    AI $\eta_2^w$ & [days] &  1676.794  & $5074 ^{+3661} _{-2669}$   & 1660.697  & $4674 ^{+3920} _{-2729}$   & 3041.504   & $5898 ^{+3313} _{-3012}$   \\[1ex] 
    RV $s$        & [m/s]  & 1.187             & $1.20 ^{+0.04} _{-0.04}$            & 1.074     & $1.129 ^{+0.056} _{-0.006}$           & 1.126      & $1.12 ^{+0.04} _{-0.04}$            \\[1ex] 
    AI $s$        & [m/s]  & 0.423             & $0.438 ^{+0.032} _{-0.030}$            & 1.229     & $1.20 ^{+0.06} _{-0.06}$            & 0.001      & $0.00100 ^{+0.00009} _{-0.00009}$              \\[1ex]  
    \hline\\
    \end{tabular}
        \caption{As in Table \ref{table:Results1}, but with respect to the GPRN
        joint analysis of the datasets, from left to right, RVs and BIS, RVs and
        FWHM, RVs and \logRhk.}
        \label{table:Results2}
\end{table*}

\subsubsection{RV and FWHM}
\begin{figure*}
    \includegraphics[width=0.9\textwidth]{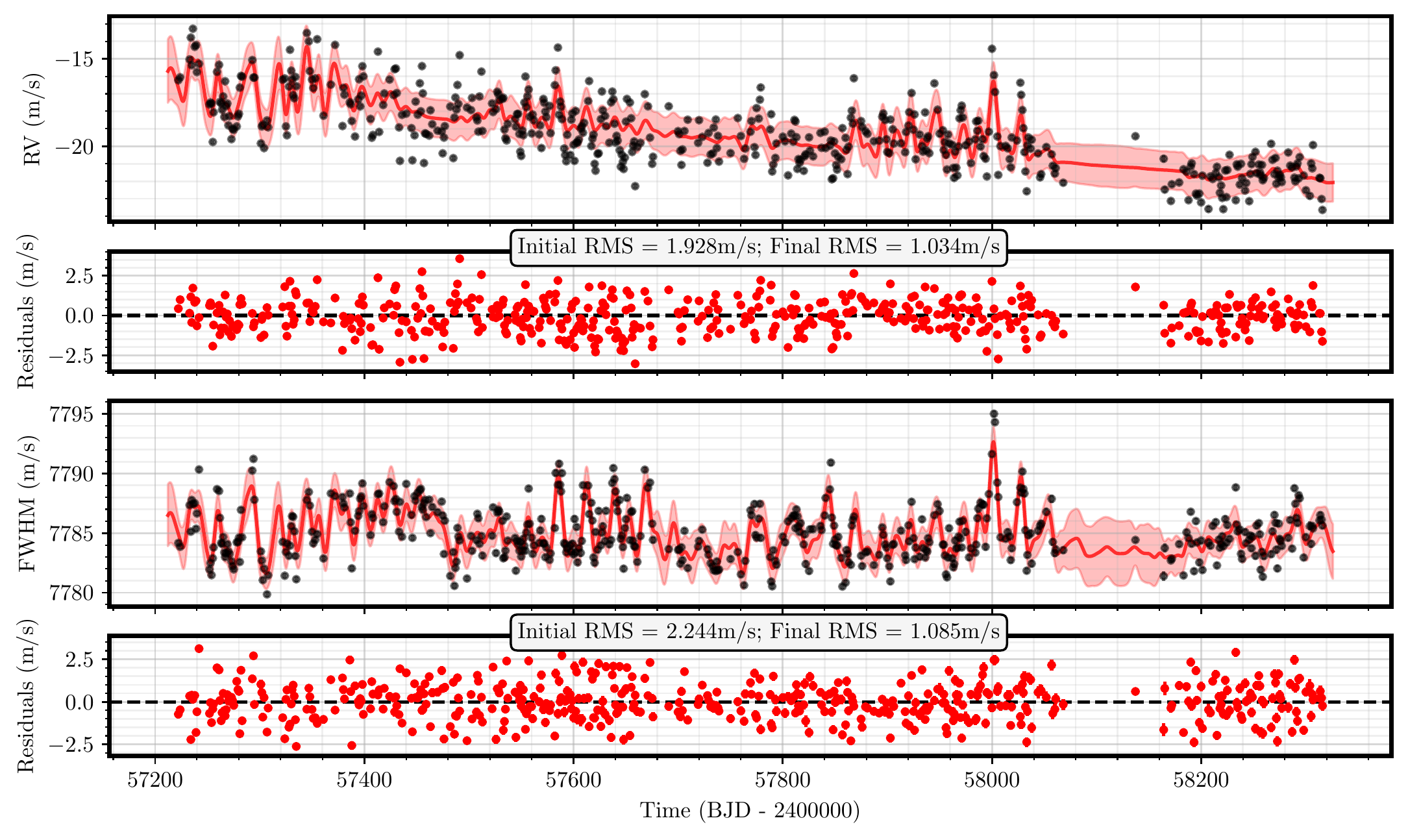}
    \caption{As in Figure \ref{fig:RVsBISfit_GPRN}, but with respect to the
     joint analysis of the RVs and FWHM datasets.}
    \label{fig:RVsFWHMfit_GPRN}
\end{figure*}

We now turn our attention to the joint analysis of the RVs and FWHM. Figure
\ref{fig:RVsFWHMfit_GPRN} shows the GPRN posterior predictive means and standard
deviations using the MAP values of the hyper-parameters. As in the previous
joint analysis, the reduction in rms, from the original datasets to the sets of
residuals obtained by subtracting the MAP predictive means from the former, is
lower than the reduction in rms obtained through the analysis of the datasets in
isolation. The RV and FWHM rms are reduced by a factor of 1.9 and 2.1, instead
of 2.8 and 2.2, respectively. Again, the node gives preference to fitting the
activity indicator. This is also reflected in the significantly lower increase
in the MAP value for the jitter, 8\% for the FWHM versus 38\% for the RVs, and
in the posterior predictive uncertainties, with respect to what was found in the
individual analysis. Lower harmonic complexity in the case of the FWHM may be
also behind this preferred behaviour for the GPRN.

We again recover a precise estimate for the synodic rotation period of the Sun,
and most of the MAP values for the other hyper-parameters are similar to what
was previously found in the joint analysis of RVs and BIS. The largest
difference is in the behaviour of the weight connecting the node to the RV
dataset, which prefers a significantly smaller timescale $\eta_2^w$. In Figure
\ref{fig:fullPlots_RVFWHM} it can be seen that indeed the preferred RV weight
function has noticeably more structure, besides a general downwards trend. 

As before, the preference of the node for fitting more closely the variations
observed in the activity indicator leaves to the weight function the task of
compensating for the node function in order to match the RV variations. But it
seems unlikely that the considerable complexity of the RV weight function is
explained solely by this effect. This suggests that the RVs and FWHM do not
share a common latent activity process to the extent that was observed for the
RVs and BIS, at least when such process is assumed to have a quasi-periodic
structure. 

\subsubsection{RV and \logRhk}
\begin{figure*}
    \includegraphics[width=0.9\textwidth]{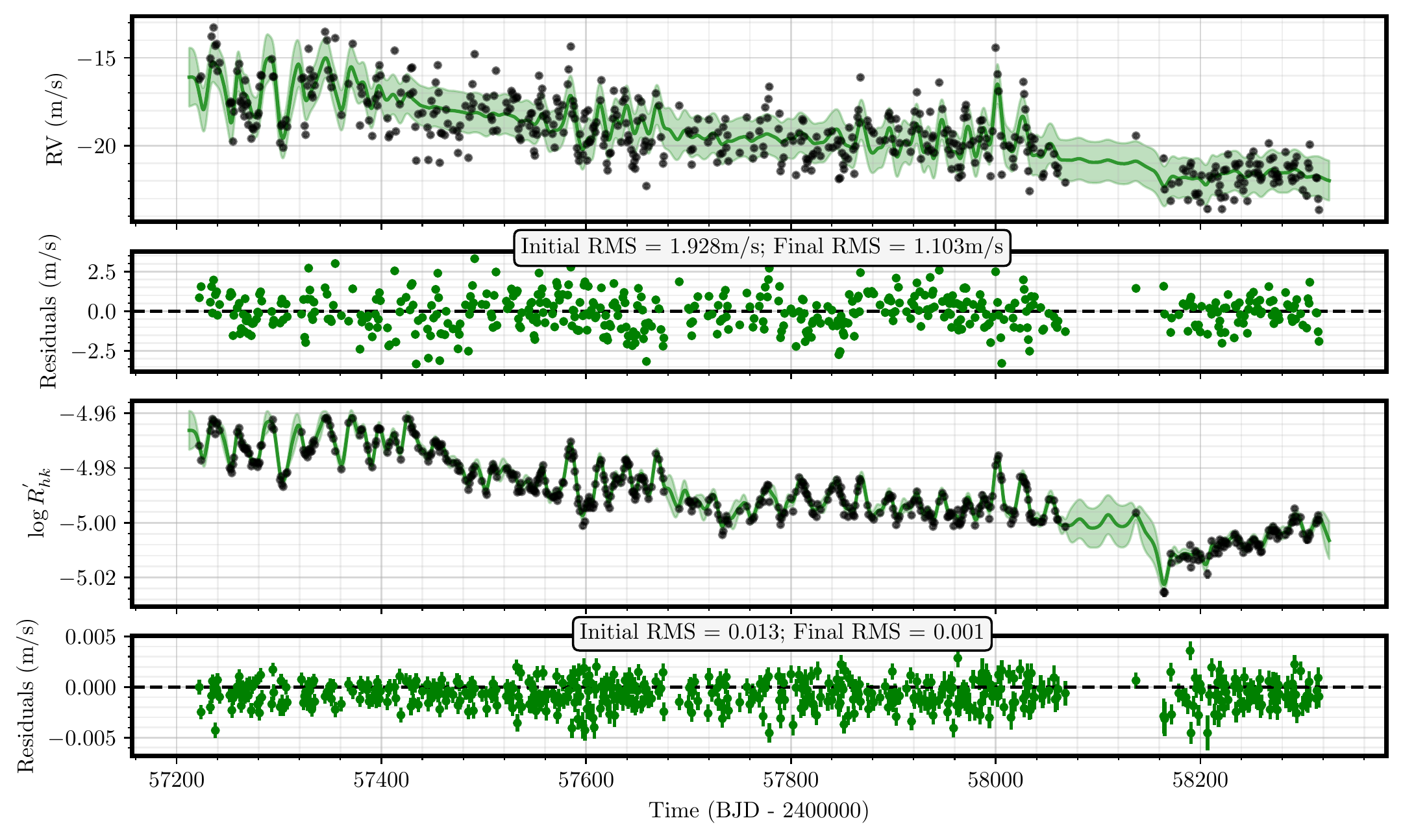}
    \caption{As in Figure \ref{fig:RVsBISfit_GPRN}, but with respect to the
    joint analysis of the RVs and \logRhk datasets.}
    \label{fig:RVsRhkfit_GPRN}
\end{figure*}

The final joint analysis considers the RV and \logRhk datasets. The GPRN
posterior predictive means and standard deviations, using the MAP values of the
hyper-parameters, are shown in Figure \ref{fig:RVsRhkfit_GPRN}. Contrary to the
previous cases, the quality of the MAP fit to the activity indicator does not
show any degradation with respect to the MAP fit obtained for the \logRhk
dataset individually. But the MAP fit to the RVs shows the typical reduction in
rms improvement when moving from the individual to the joint analysis,
decreasing from a factor of 2.8 to about 1.7. 

The \logRhk periodogram resembles best what one would expect from a quasi-periodic 
process centred around a dominant (stellar rotation) period 
(see Figure \ref{fig:sunMeasurements}), as assumed in the set-up of the GPRN. Thus, 
it should not be surprising that the joint analysis leads to a preferred model 
configuration where the node essentially fits the \logRhk, leaving for the RV weight 
(for which the $\eta_2^w$ MAP is again relatively small, around 88 days) the onus of
adapting the node output to match the RV dataset (see Figure
\ref{fig:fullPlots_RVRHK}). 

\subsection{Interpretation of the results}

\begin{figure}
    \includegraphics[width=0.45\textwidth]{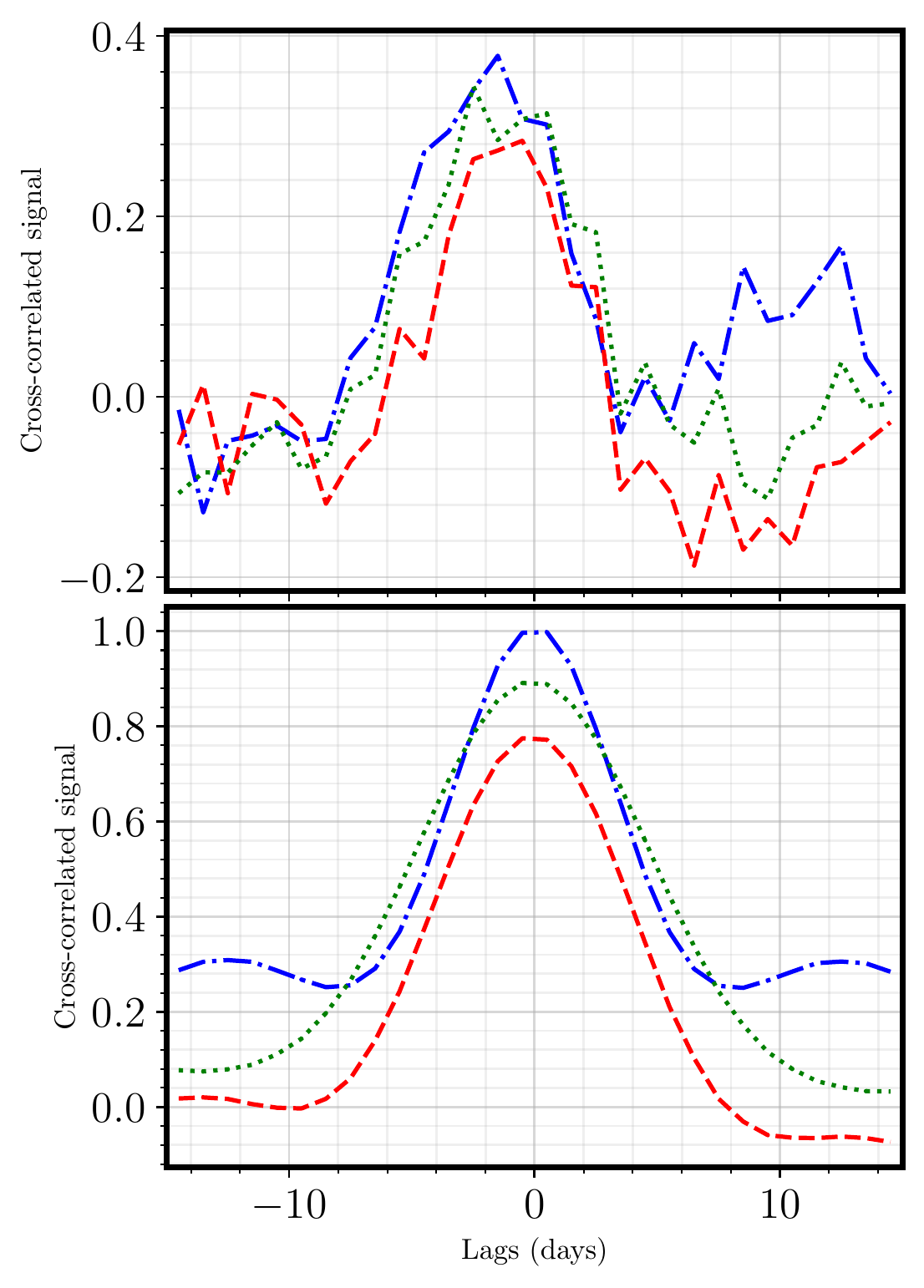}
    \caption{Cross-correlation functions (CCFs) between the RVs and activity
    indicators - BIS (in blue), FWHM (in red) and $\log R_{hk}^{'}$ (in green).
    The upper panel shows the CCFs between the observed time series, while the
    lower panel shows the CCFs between the MAP predictive means at the observed
    timestamps, obtained through the joint analysis of the measured RVs plus the
    given activity indicator.}
    \label{fig:lags}
\end{figure}

We have already seen that we can recover the synodic solar rotation period when
the time series are analysed both individually and jointly. In fact, overall the
MAP values obtained for the GPRN hyper-parameters in all analyses are very
similar (see Tables \ref{table:Results1} and \ref{table:Results2}). This
suggests the RVs and activity indicators share the same type of temporal
evolution. 

The GPRN model we considered is able to describe very well the behaviour of each
time series individually, leading to substantial reductions in rms and in the
significance of periodogram peaks when the MAP predictive means are subtracted
from the original datasets (see Figure \ref{fig:residuals_individual}). However,
such reductions are significantly smaller for the RVs when this dataset is
analysed jointly with an activity indicator. A degradation in the quality of the
best-fit solutions was expected to some extent given that in the joint analysis
we are forcing the node to model two independently acquired datasets.
Nevertheless, the magnitude of such degradation in the case of the RVs is
surprising given how close the MAP values obtained for the GPRN hyper-parameters
are for all individual analysis. One possible cause for this may be the presence
of temporal delays between the solar RVs and activity indicators, which the GPRN
model, with only one node, should find very difficult to reproduce.

Indeed, time lags of 1 and 3 days between the RVs and the FWHM and BIS
measurements, respectively, were identified by \cite{Cameron2019_SunAsAStar}.
Since the datasets we consider differ somewhat from the ones they assembled (see
section \ref{ch:Sun}), we used the algorithm developed by
\cite{Edeldon1988_discreteCCF} to determine the discrete correlation function
between the RVs and activity indicators, in search for possible temporal delays
in our datasets. The results are shown in the upper panel of Figure
\ref{fig:lags}, and suggest that the RVs lead all activity indicators by 1 to 2
days. As expected, the MAP predictive means of the RVs and activity indicators
obtained in the joint analysis do not show any sign of such time lags, as can be
seen through their correlation functions in the lower panel of Figure
\ref{fig:lags}. The failure to model this aspect of the cross-correlation
between the time series should lead to worse fits when they are jointly
modelled, as found.

Finally, both the individual and joint analysis of the time series lead to
preferred posterior predicted means for the weight functions that decrease over
the timespan of the observations. This implies an overall decrease in the
variances of the time series, correctly reproducing the expected diminishing
solar activity as the Sun approached the minimum of its 11-years sunspot cycle.
Thus, the detection and characterisation of the non-stationary behaviour of the
activity indicators does not seem to be much impacted by the inability of the
one-node GPRN to model the time lag between the RVs and activity indicators.

\section{Discussion and conclusions}
\label{ch:DiscussionConclusions}

We have presented a new framework to model RV observations in combination with
activity indicators. A GPRN combines $Q$ nodes and $Q\times P$ weights to model
$P$ time series, with all nodes and weights defined as independent GPs. This
makes it possible to model non-stationary outputs, an improvement with respect
to currently existing frameworks.

We tested our GPRN implementation on simulated datasets, and demonstrated its
capabilities on real Sun-as-a-star observations obtained with the HARPS-N
spectrograph. We obtained physically sensible values for the GPRN
hyper-parameters, recovering for example the synodic solar rotation period, both
when the time series were analysed individually and jointly.

The joint analysis of the RVs and an activity indicator leads to a
significantly worse description of the RV variations with time than when the RVs
are analysed in isolation. This is unfortunate, because such joint analysis may
help disentangle better the RV components due to stellar activity and orbiting
planets. However, we only considered the simplest possible configuration for a
GPRN, with just one node. The inclusion of extra nodes would help in modelling more
complex behaviour, in particular that which is not shared by the outputs being
analysed. 

Previous work by \citet{Rajpaul2015_GPFramework} and
\citet{Gilbertson2020_JonesFramework}, for example, proposed to use a linear
combination of a GP and its derivatives to model the RV variations induced by
spots, due to their dependence on both the spot coverage of the stellar surface
and the suppression of convective blueshift in magnetised regions. Purportedly,
the BIS shares the same dependencies as the RVs, but the FWHM and \logRhk would
be sensitive only to the fraction of the stellar surface covered by spots and,
as such, would not require the use of the derivative term. Our results seem to
lend some support to this hypothesis, given that the node by itself seems to be
able to describe better the RVs jointly with the BIS than with either the FWHM
or the $\log R_{hk}^{'}$. This can be perceived through the significantly higher
complexity required for the weight function associated with the RVs in the later
cases (cf. Figure~\ref{fig:fullPlots_RVBIS} with
Figures~\ref{fig:fullPlots_RVFWHM} and~\ref{fig:fullPlots_RVRHK}). 

On the other hand, the MAP fit to the RVs when modelled jointly with the BIS
leads to a worse residual rms then what is obtained in the other joint analyses.
These contradicting results could be due to the seemingly larger time lag
between the RVs and the BIS than with respect to the other activity indicators
(see Figure \ref{fig:lags}). This behaviour is difficult to model by the weight
associated with the RVs and could justify the worse residual rms when the RVs
are analysed jointly with the BIS as well as push such weight towards the
adoption of a simpler form.

The joint analysis of the RVs and an activity indicator could thus benefit from
the inclusion of a second node in the GPRN with a kernel that is the derivative
of the kernel assumed for the first node. In particular, if the later is
periodic or quasi-periodic, the association of that kernel derivative with a
second node would allow for the modelling of any time lag that may exist between
the outputs. However, such feature could eventually be also modelled directly,
through the introduction of $P-1$ lag parameters, one per output, while keeping
just one node. 

Nevertheless, if the main objective is to use the information in activity
indicators to constrain the behaviour of the RV component induced by stellar
activity, care must be taken not to increase the flexibility of the GPRN to the
point where it could model RVs and activity indicator(s) almost independently.
While the probability of this happening increases with the number of nodes
assumed, it can be mitigated to some extent by not allowing the number of 
hyper-parameters to grow proportionally.

Another aspect that we would like to study in the future pertains to the type
and combination of kernels used in the nodes and weights, including the
introduction of non-stationary white noise. In particular, we cannot exclude the
possibility that our choice of a squared exponential kernel for the weights
limits too much the extent to which the node output can be modified. Although
widely used, this kernel is known to be too smooth to realistically model a
considerable number of physical processes
\citep[e.g.][]{Stein1999_MaternKernels}.

Interestingly, although the simple GPRN model we considered seems unable to
fully model RVs and activity indicators jointly, it is still capable of
describing the RVs well enough for the rms to be reduced by a factor of 1.6 to
1.9 ($1.93$ m/s $\rightarrow$ $1.18$ - $1.03$ m/s) when the MAP predictive means
are subtracted from the RV measurements. These are higher rms reduction factors
than what has been previously reported for HARPS-N solar RVs in the published
literature. For example, \cite{Langellier2020_detectionLimitsGPs} achieved a rms
reduction factor of 1.4, from $1.65$ m/s to $1.14$ m/s, by analysing HARPS-N
solar RVs with a GP where all the quasi-periodic kernel hyper-parameters, except
the amplitude, were fixed a priori to the MAP values obtained through a previous
GP-based analysis of the HARPS-N solar S-index data. The same rms reduction
factor was also obtained by \cite{Cameron2021_scalpels}, from $1.76$ m/s to
$1.25$ m/s, and \cite{Milbourne2019_SolarRVs}, from $1.65$ m/s to $1.21$ m/s,
using very different methodologies. Although some of the rms reduction 
factors mentioned above were obtained without taking into account a linear trend, our results are almost unchanged when we do not include such a trend in the mean function of the GPRN.
Further, given that the GPRN posterior predictive distribution has heavier-tails 
than a Gaussian, the quality of the best-fit GPRN model to the HARPS-N solar data 
may actually be better than what is suggested by the residual rms.

Ultimately, the usefulness of the methods that have been proposed to help
disentangle the RV contributions due to stellar activity and orbiting planets
will have to be tested and compared using data for stars other than the Sun. The
inclusion of Keplerians in the mean function associated with the RVs is the only
change required in the GPRN framework we presented for it to be ready for
such purpose.

\section*{Acknowledgements}
\label{ch:AckAck}
We thank Nuno C. Santos for the many valuable discussions and continuous support
during the development of this work. This work was supported by FCT, Fundação
para a Ciência e a Tecnologia, through national funds and by FEDER through
COMPETE2020, Programa Operacional Competitividade e Internacionalização, by the
following grants: UID/FIS/04434/2019; UIDB/04434/2020; UIDP/04434/2020;
PTDC/FIS-AST/32113/2017 and POCI-01-0145-FEDER-032113; PTDC/FIS-AST/28953/2017
and POCI-01-0145-FEDER-028953; PTDC/FIS-AST/28987/2017 and
POCI-01-0145-FEDER-028987; PTDC/FIS-AST/30389/2017 and
POCI-01-0145-FEDER-030389; EXPL/FIS-AST/0615/2021. 
J.P.F. is supported in the form of a work contract funded by national funds
through FCT with reference DL57/2016/CP1364/CT0005.
\section*{Data Availability}
\label{ch:aAckAck}
The data and software used in this article are publicly available online in \texttt{gpyrn} GitHub repository\footnote{~\url{https://github.com/iastro-pt/gpyrn}}.

\bibliographystyle{mnras}
\bibliography{MyBibliography.bib}

\begin{thebibliography}{}
\makeatletter
\relax
\def\mn@urlcharsother{\let\do\@makeother \do\$\do\&\do\#\do\^\do\_\do\%\do\~}
\def\mn@doi{\begingroup\mn@urlcharsother \@ifnextchar [ {\mn@doi@}
  {\mn@doi@[]}}
\def\mn@doi@[#1]#2{\def\@tempa{#1}\ifx\@tempa\@empty \href
  {http://dx.doi.org/#2} {doi:#2}\else \href {http://dx.doi.org/#2} {#1}\fi
  \endgroup}
\def\mn@eprint#1#2{\mn@eprint@#1:#2::\@nil}
\def\mn@eprint@arXiv#1{\href {http://arxiv.org/abs/#1} {{\tt arXiv:#1}}}
\def\mn@eprint@dblp#1{\href {http://dblp.uni-trier.de/rec/bibtex/#1.xml}
  {dblp:#1}}
\def\mn@eprint@#1:#2:#3:#4\@nil{\def\@tempa {#1}\def\@tempb {#2}\def\@tempc
  {#3}\ifx \@tempc \@empty \let \@tempc \@tempb \let \@tempb \@tempa \fi \ifx
  \@tempb \@empty \def\@tempb {arXiv}\fi \@ifundefined
  {mn@eprint@\@tempb}{\@tempb:\@tempc}{\expandafter \expandafter \csname
  mn@eprint@\@tempb\endcsname \expandafter{\@tempc}}}

\bibitem[\protect\citeauthoryear{{Aigrain}, {Pont}  \& {Zucker}}{{Aigrain}
  et~al.}{2012}]{Aigrain2012_FFMethod}
{Aigrain} S.,  {Pont} F.,   {Zucker} S.,  2012, \mn@doi [\mnras]
  {10.1111/j.1365-2966.2011.19960.x}, \href
  {https://ui.adsabs.harvard.edu/abs/2012MNRAS.419.3147A} {419, 3147}

\bibitem[\protect\citeauthoryear{Arfken, Weber  \& Harris}{Arfken
  et~al.}{2013}]{Arfken2013_Bessel}
Arfken G.,  Weber H.,   Harris F.,  2013, Mathematical Methods for Physicists:
  A Comprehensive Guide.
Elsevier Science

\bibitem[\protect\citeauthoryear{{Astropy Collaboration} et~al.,}{{Astropy
  Collaboration} et~al.}{2018}]{Astropy2018}
{Astropy Collaboration} et~al., 2018, \mn@doi [\aj] {10.3847/1538-3881/aabc4f},
  \href {https://ui.adsabs.harvard.edu/abs/2018AJ....156..123A} {156, 123}

\bibitem[\protect\citeauthoryear{{Baliunas} et~al.,}{{Baliunas}
  et~al.}{1995}]{Baliunas1995_magCycles}
{Baliunas} S.~L.,  et~al., 1995, \mn@doi [\apj] {10.1086/175072}, \href
  {https://ui.adsabs.harvard.edu/abs/1995ApJ...438..269B} {438, 269}

\bibitem[\protect\citeauthoryear{{Barrag{\'a}n}, {Gandolfi}  \&
  {Antoniciello}}{{Barrag{\'a}n} et~al.}{2019a}]{Barragan2019_pyaneti1}
{Barrag{\'a}n} O.,  {Gandolfi} D.,   {Antoniciello} G.,  2019a, \mn@doi
  [\mnras] {10.1093/mnras/sty2472}, \href
  {https://ui.adsabs.harvard.edu/abs/2019MNRAS.482.1017B} {482, 1017}

\bibitem[\protect\citeauthoryear{{Barrag{\'a}n} et~al.,}{{Barrag{\'a}n}
  et~al.}{2019b}]{Barragan2019_K2100b}
{Barrag{\'a}n} O.,  et~al., 2019b, \mn@doi [\mnras] {10.1093/mnras/stz2569},
  \href {https://ui.adsabs.harvard.edu/abs/2019MNRAS.490..698B} {490, 698}

\bibitem[\protect\citeauthoryear{{Barrag{\'a}n}, {Aigrain}, {Rajpaul}  \&
  {Zicher}}{{Barrag{\'a}n} et~al.}{2021}]{Barragan2021_pyaneti2}
{Barrag{\'a}n} O.,  {Aigrain} S.,  {Rajpaul} V.~M.,   {Zicher} N.,  2021,
  \mn@doi [\mnras] {10.1093/mnras/stab2889}, \href
  {https://ui.adsabs.harvard.edu/abs/2021MNRAS.tmp.2625B} {509, 866}

\bibitem[\protect\citeauthoryear{{Bazot}, {Bouchy}, {Kjeldsen}, {Charpinet},
  {Laymand}  \& {Vauclair}}{{Bazot} et~al.}{2007}]{Bazot2007_oscillations}
{Bazot} M.,  {Bouchy} F.,  {Kjeldsen} H.,  {Charpinet} S.,  {Laymand} M.,
  {Vauclair} S.,  2007, \mn@doi [\aap] {10.1051/0004-6361:20065694}, \href
  {https://ui.adsabs.harvard.edu/abs/2007A&A...470..295B} {470, 295}

\bibitem[\protect\citeauthoryear{Blei, Kucukelbir  \& McAuliffe}{Blei
  et~al.}{2017}]{Blei2018_VarInference}
Blei D.~M.,  Kucukelbir A.,   McAuliffe J.~D.,  2017, \mn@doi [Journal of the
  American Statistical Association] {10.1080/01621459.2017.1285773}, 112, 859

\bibitem[\protect\citeauthoryear{{Cegla} et~al.,}{{Cegla}
  et~al.}{2012}]{Cegla2012_GravRedshift}
{Cegla} H.~M.,  et~al., 2012, \mn@doi [\mnras]
  {10.1111/j.1745-3933.2011.01205.x}, \href
  {https://ui.adsabs.harvard.edu/abs/2012MNRAS.421L..54C} {421, L54}

\bibitem[\protect\citeauthoryear{{Charbonneau}}{{Charbonneau}}{2014}]{Charbonneau2014_solarDynamo}
{Charbonneau} P.,  2014, \mn@doi [\araa] {10.1146/annurev-astro-081913-040012},
  \href {https://ui.adsabs.harvard.edu/abs/2014ARA&A..52..251C} {52, 251}

\bibitem[\protect\citeauthoryear{{Cloutier} et~al.,}{{Cloutier}
  et~al.}{2019}]{Cloutier2019_K218}
{Cloutier} R.,  et~al., 2019, \mn@doi [\aap] {10.1051/0004-6361/201833995},
  \href {https://ui.adsabs.harvard.edu/abs/2019A&A...621A..49C} {621, A49}

\bibitem[\protect\citeauthoryear{{Collier Cameron} et~al.,}{{Collier Cameron}
  et~al.}{2019}]{Cameron2019_SunAsAStar}
{Collier Cameron} A.,  et~al., 2019, \mn@doi [\mnras] {10.1093/mnras/stz1215},
  \href {https://ui.adsabs.harvard.edu/abs/2019MNRAS.487.1082C} {487, 1082}

\bibitem[\protect\citeauthoryear{{Collier Cameron} et~al.,}{{Collier Cameron}
  et~al.}{2021}]{Cameron2021_scalpels}
{Collier Cameron} A.,  et~al., 2021, \mn@doi [\mnras] {10.1093/mnras/stab1323},
  \href {https://ui.adsabs.harvard.edu/abs/2021MNRAS.505.1699C} {505, 1699}

\bibitem[\protect\citeauthoryear{{Cui}, {Yu}, {Iommelli}  \& {Kong}}{{Cui}
  et~al.}{2016}]{Cui2016_Bessel}
{Cui} G.,  {Yu} X.,  {Iommelli} S.,   {Kong} L.,  2016, \mn@doi [IEEE Signal
  Processing Letters] {10.1109/LSP.2016.2614539}, \href
  {https://ui.adsabs.harvard.edu/abs/2016ISPL...23.1662C} {23, 1662}

\bibitem[\protect\citeauthoryear{{Del Zanna} \& {Mason}}{{Del Zanna} \&
  {Mason}}{2013}]{DelZanna2013_sunInterior}
{Del Zanna} G.,  {Mason} H.,  2013, in Planets, Stars and Stellar Systems:
  Volume 4: Stellar Structure and Evolution. Springer Netherlands, Dordrecht,
  pp 87--205

\bibitem[\protect\citeauthoryear{{Delisle}, {Unger}, {Hara}  \&
  {S{\'e}gransan}}{{Delisle} et~al.}{2022}]{Delisle2022}
{Delisle} J.~B.,  {Unger} N.,  {Hara} N.~C.,   {S{\'e}gransan} D.,  2022,
  \mn@doi [\aap] {10.1051/0004-6361/202141949}, \href
  {https://ui.adsabs.harvard.edu/abs/2022A&A...659A.182D} {659, A182}

\bibitem[\protect\citeauthoryear{{Demin}, {Nefedyev}, {Andreev}, {Demina}  \&
  {Timashev}}{{Demin} et~al.}{2018}]{Demin2018_NonStatSolarActivity}
{Demin} S.~A.,  {Nefedyev} Y.~A.,  {Andreev} A.~O.,  {Demina} N.~Y.,
  {Timashev} S.~F.,  2018, \mn@doi [Advances in Space Research]
  {10.1016/j.asr.2017.06.055}, \href
  {https://ui.adsabs.harvard.edu/abs/2018AdSpR..61..639D} {61, 639}

\bibitem[\protect\citeauthoryear{{Dumusque}, {Udry}, {Lovis}, {Santos}  \&
  {Monteiro}}{{Dumusque} et~al.}{2011}]{Dumusque2011a}
{Dumusque} X.,  {Udry} S.,  {Lovis} C.,  {Santos} N.~C.,   {Monteiro}
  M.~J.~P.~F.~G.,  2011, \mn@doi [\aap] {10.1051/0004-6361/201014097}, \href
  {https://ui.adsabs.harvard.edu/abs/2011A&A...525A.140D} {525, A140}

\bibitem[\protect\citeauthoryear{{Dumusque} et~al.,}{{Dumusque}
  et~al.}{2015}]{Dumusque2015_HARPS-N}
{Dumusque} X.,  et~al., 2015, \mn@doi [\apjl] {10.1088/2041-8205/814/2/L21},
  \href {https://ui.adsabs.harvard.edu/abs/2015ApJ...814L..21D} {814, L21}

\bibitem[\protect\citeauthoryear{{Dumusque} et~al.,}{{Dumusque}
  et~al.}{2021}]{Dumusque2021_SunRVs}
{Dumusque} X.,  et~al., 2021, \mn@doi [\aap] {10.1051/0004-6361/202039350},
  \href {https://ui.adsabs.harvard.edu/abs/2021A&A...648A.103D} {648, A103}

\bibitem[\protect\citeauthoryear{{Edelson} \& {Krolik}}{{Edelson} \&
  {Krolik}}{1988}]{Edeldon1988_discreteCCF}
{Edelson} R.~A.,  {Krolik} J.~H.,  1988, \mn@doi [\apj] {10.1086/166773}, \href
  {https://ui.adsabs.harvard.edu/abs/1988ApJ...333..646E} {333, 646}

\bibitem[\protect\citeauthoryear{{Faria}, {Haywood}, {Brewer}, {Figueira},
  {Oshagh}, {Santerne}  \& {Santos}}{{Faria} et~al.}{2016}]{Faria2016_Corot7}
{Faria} J.~P.,  {Haywood} R.~D.,  {Brewer} B.~J.,  {Figueira} P.,  {Oshagh} M.,
   {Santerne} A.,   {Santos} N.~C.,  2016, \mn@doi [\aap]
  {10.1051/0004-6361/201527899}, \href
  {https://ui.adsabs.harvard.edu/abs/2016A&A...588A..31F} {588, A31}

\bibitem[\protect\citeauthoryear{{Figueira} et~al.,}{{Figueira}
  et~al.}{2010}]{Figueira2010_BD201790}
{Figueira} P.,  et~al., 2010, \mn@doi [\aap] {10.1051/0004-6361/201014323},
  \href {https://ui.adsabs.harvard.edu/abs/2010A&A...513L...8F} {513, L8}

\bibitem[\protect\citeauthoryear{{Figueira}, {Santos}, {Pepe}, {Lovis}  \&
  {Nardetto}}{{Figueira} et~al.}{2013}]{Figueira2013_Indicators}
{Figueira} P.,  {Santos} N.~C.,  {Pepe} F.,  {Lovis} C.,   {Nardetto} N.,
  2013, \mn@doi [\aap] {10.1051/0004-6361/201220779}, \href
  {https://ui.adsabs.harvard.edu/abs/2013A&A...557A..93F} {557, A93}

\bibitem[\protect\citeauthoryear{{Foreman-Mackey}, {Agol}, {Ambikasaran}  \&
  {Angus}}{{Foreman-Mackey} et~al.}{2017}]{celerite}
{Foreman-Mackey} D.,  {Agol} E.,  {Ambikasaran} S.,   {Angus} R.,  2017,
  \mn@doi [\aj] {10.3847/1538-3881/aa9332}, \href
  {https://ui.adsabs.harvard.edu/abs/2017AJ....154..220F} {154, 220}

\bibitem[\protect\citeauthoryear{{Foreman-Mackey} et~al.,}{{Foreman-Mackey}
  et~al.}{2019}]{ForemanMackey2019_emcee}
{Foreman-Mackey} D.,  et~al., 2019, \mn@doi [JOSS] {10.21105/joss.01864}, \href
  {https://ui.adsabs.harvard.edu/abs/2019JOSS....4.1864F} {4, 1864}

\bibitem[\protect\citeauthoryear{{Gilbertson}, {Ford}, {Jones}  \&
  {Stenning}}{{Gilbertson} et~al.}{2020}]{Gilbertson2020_JonesFramework}
{Gilbertson} C.,  {Ford} E.~B.,  {Jones} D.~E.,   {Stenning} D.~C.,  2020,
  \mn@doi [\apj] {10.3847/1538-4357/abc627}, \href
  {https://ui.adsabs.harvard.edu/abs/2020ApJ...905..155G} {905, 155}

\bibitem[\protect\citeauthoryear{{Gomes da Silva}, {Santos}, {Bonfils},
  {Delfosse}, {Forveille}, {Udry}, {Dumusque}  \& {Lovis}}{{Gomes da Silva}
  et~al.}{2012}]{Gomesdasilva2012_Activity}
{Gomes da Silva} J.,  {Santos} N.~C.,  {Bonfils} X.,  {Delfosse} X.,
  {Forveille} T.,  {Udry} S.,  {Dumusque} X.,   {Lovis} C.,  2012, \mn@doi
  [\aap] {10.1051/0004-6361/201118598}, \href
  {https://ui.adsabs.harvard.edu/abs/2012A&A...541A...9G} {541, A9}

\bibitem[\protect\citeauthoryear{{Gregory}}{{Gregory}}{2005}]{Gregory2005}
{Gregory} P.~C.,  2005, \mn@doi [\apj] {10.1086/432594}, \href
  {https://ui.adsabs.harvard.edu/abs/2005ApJ...631.1198G} {631, 1198}

\bibitem[\protect\citeauthoryear{{Guinan}, {Engle}  \& {Durbin}}{{Guinan}
  et~al.}{2016}]{Guinan2016_KapteynStar}
{Guinan} E.~F.,  {Engle} S.~G.,   {Durbin} A.,  2016, \mn@doi [\apj]
  {10.3847/0004-637X/821/2/81}, \href
  {https://ui.adsabs.harvard.edu/abs/2016ApJ...821...81G} {821, 81}

\bibitem[\protect\citeauthoryear{{Hathaway}}{{Hathaway}}{2015}]{Hathaway2015_solarCycle}
{Hathaway} D.~H.,  2015, \mn@doi [Living Reviews in Solar Physics]
  {10.1007/lrsp-2015-4}, \href
  {https://ui.adsabs.harvard.edu/abs/2015LRSP...12....4H} {12, 4}

\bibitem[\protect\citeauthoryear{{Hathaway} \& {Choudhary}}{{Hathaway} \&
  {Choudhary}}{2008}]{Hathaway2008_sunspotDecay}
{Hathaway} D.~H.,  {Choudhary} D.~P.,  2008, \mn@doi [\solphys]
  {10.1007/s11207-008-9226-4}, \href
  {https://ui.adsabs.harvard.edu/abs/2008SoPh..250..269H} {250, 269}

\bibitem[\protect\citeauthoryear{{Haywood} et~al.,}{{Haywood}
  et~al.}{2014}]{Haywood2014_Corot7}
{Haywood} R.~D.,  et~al., 2014, \mn@doi [\mnras] {10.1093/mnras/stu1320}, \href
  {https://ui.adsabs.harvard.edu/abs/2014MNRAS.443.2517H} {443, 2517}

\bibitem[\protect\citeauthoryear{Heinonen, Mannerström, Rousu, Kaski  \&
  Lähdesmäki}{Heinonen et~al.}{2016}]{Heinonen2015_nonStationaryGPs}
Heinonen M.,  Mannerström H.,  Rousu J.,  Kaski S.,   Lähdesmäki H.,  2016,
  in Gretton A.,  Robert C.~C.,  eds,  Proceedings of Machine Learning Research
  Vol. 51, Proceedings of the 19th International Conference on Artificial
  Intelligence and Statistics. PMLR, Cadiz, Spain, pp 732--740

\bibitem[\protect\citeauthoryear{{Jones}, {Stenning}, {Ford}, {Wolpert},
  {Loredo}  \& {Dumusque}}{{Jones} et~al.}{2017}]{Jones2017_GPFramework}
{Jones} D.~E.,  {Stenning} D.~C.,  {Ford} E.~B.,  {Wolpert} R.~L.,  {Loredo}
  T.~J.,   {Dumusque} X.,  2017, preprint (arXiv:1711.01318)

\bibitem[\protect\citeauthoryear{Jordan, Ghahramani, Jaakkola  \& Saul}{Jordan
  et~al.}{1999}]{Jordan1999_variationalInference}
Jordan M.~I.,  Ghahramani Z.,  Jaakkola T.~S.,   Saul L.~K.,  1999, Machine
  learning, 37, 183

\bibitem[\protect\citeauthoryear{{Karamanis}, {Beutler}  \&
  {Peacock}}{{Karamanis} et~al.}{2021}]{Karamanis2021_zeus}
{Karamanis} M.,  {Beutler} F.,   {Peacock} J.~A.,  2021, \mn@doi [\mnras]
  {10.1093/mnras/stab2867}, \href
  {https://ui.adsabs.harvard.edu/abs/2021MNRAS.508.3589K} {508, 3589}

\bibitem[\protect\citeauthoryear{{Kjeldsen} et~al.,}{{Kjeldsen}
  et~al.}{2008}]{Kjeldsen2008_oscillations}
{Kjeldsen} H.,  et~al., 2008, \mn@doi [\apj] {10.1086/589142}, \href
  {https://ui.adsabs.harvard.edu/abs/2008ApJ...682.1370K} {682, 1370}

\bibitem[\protect\citeauthoryear{{Kosiarek} \& {Crossfield}}{{Kosiarek} \&
  {Crossfield}}{2020}]{Kosiarek2020_photometryAsProxy}
{Kosiarek} M.~R.,  {Crossfield} I. J.~M.,  2020, \mn@doi [\aj]
  {10.3847/1538-3881/ab8d3a}, \href
  {https://ui.adsabs.harvard.edu/abs/2020AJ....159..271K} {159, 271}

\bibitem[\protect\citeauthoryear{Kullback}{Kullback}{1959}]{Kullback59}
Kullback S.,  1959, Information Theory and Statistics.
Wiley, New York

\bibitem[\protect\citeauthoryear{{Lagrange}, {Desort}  \& {Meunier}}{{Lagrange}
  et~al.}{2010}]{Lagrange2010_ColdSpots}
{Lagrange} A.~M.,  {Desort} M.,   {Meunier} N.,  2010, \mn@doi [\aap]
  {10.1051/0004-6361/200913071}, \href
  {https://ui.adsabs.harvard.edu/abs/2010A&A...512A..38L} {512, A38}

\bibitem[\protect\citeauthoryear{{Langellier} et~al.,}{{Langellier}
  et~al.}{2021}]{Langellier2020_detectionLimitsGPs}
{Langellier} N.,  et~al., 2021, \mn@doi [\aj] {10.3847/1538-3881/abf1e0}, \href
  {https://ui.adsabs.harvard.edu/abs/2021AJ....161..287L} {161, 287}

\bibitem[\protect\citeauthoryear{Li, Xing, Kirby  \& Zhe}{Li
  et~al.}{2020}]{li2020_gprn}
Li S.,  Xing W.,  Kirby R.~M.,   Zhe S.,  2020, in Proceedings of the
  Twenty-Ninth International Joint Conference on Artificial Intelligence,
  {IJCAI-20}. International Joint Conferences on Artificial Intelligence
  Organization, pp 2456--2462

\bibitem[\protect\citeauthoryear{{Mayo} et~al.,}{{Mayo}
  et~al.}{2019}]{Mayo2019_Kepler538b}
{Mayo} A.~W.,  et~al., 2019, \mn@doi [\aj] {10.3847/1538-3881/ab3e2f}, \href
  {https://ui.adsabs.harvard.edu/abs/2019AJ....158..165M} {158, 165}

\bibitem[\protect\citeauthoryear{{Mayor} \& {Queloz}}{{Mayor} \&
  {Queloz}}{1995}]{Mayor1995}
{Mayor} M.,  {Queloz} D.,  1995, \mn@doi [\nat] {10.1038/378355a0}, \href
  {https://ui.adsabs.harvard.edu/abs/1995Natur.378..355M} {378, 355}

\bibitem[\protect\citeauthoryear{{Meunier} \& {Lagrange}}{{Meunier} \&
  {Lagrange}}{2019}]{Meunier2019_SuperGranulation}
{Meunier} N.,  {Lagrange} A.~M.,  2019, \mn@doi [\aap]
  {10.1051/0004-6361/201935099}, \href
  {https://ui.adsabs.harvard.edu/abs/2019A&A...625L...6M} {625, L6}

\bibitem[\protect\citeauthoryear{{Milbourne} et~al.,}{{Milbourne}
  et~al.}{2019}]{Milbourne2019_SolarRVs}
{Milbourne} T.~W.,  et~al., 2019, \mn@doi [\apj] {10.3847/1538-4357/ab064a},
  \href {http://adsabs.harvard.edu/abs/2019arXiv190204184M} {874, 107}

\bibitem[\protect\citeauthoryear{Murray, Adams  \& MacKay}{Murray
  et~al.}{2010}]{murray2010_ESS}
Murray I.,  Adams R.~P.,   MacKay D. J.~C.,  2010, in The Proceedings of the
  13th International Conference on Artificial Intelligence and Statistics. pp
  541--548

\bibitem[\protect\citeauthoryear{Nguyen}{Nguyen}{2015}]{Nguyen2015_thesis}
Nguyen T.~V.,  2015, PhD thesis, Australian National University, Research
  School of Computer Science

\bibitem[\protect\citeauthoryear{Nguyen \& Bonilla}{Nguyen \&
  Bonilla}{2013}]{Nguyen2013b_GPRN}
Nguyen T.,  Bonilla E.,  2013, in Proceedings of the Sixteenth International
  Conference on Artificial Intelligence and Statistics. PMLR, Scottsdale,
  Arizona, USA, pp 472--480

\bibitem[\protect\citeauthoryear{Pagano}{Pagano}{2013}]{Pagano2013_stellarActivity}
Pagano I.,  2013, in Planets, Stars and Stellar Systems: Volume 4: Stellar
  Structure and Evolution. Springer Netherlands, Dordrecht, pp 485--557

\bibitem[\protect\citeauthoryear{{Pepe} et~al.,}{{Pepe}
  et~al.}{2021}]{Pepe2021_ESPRESSO_DRS}
{Pepe} F.,  et~al., 2021, \mn@doi [\aap] {10.1051/0004-6361/202038306}, \href
  {https://ui.adsabs.harvard.edu/abs/2021A&A...645A..96P} {645, A96}

\bibitem[\protect\citeauthoryear{{Petersburg} et~al.,}{{Petersburg}
  et~al.}{2020}]{Petersburg2020_EXPRES}
{Petersburg} R.~R.,  et~al., 2020, \mn@doi [\aj] {10.3847/1538-3881/ab7e31},
  \href {https://ui.adsabs.harvard.edu/abs/2020AJ....159..187P} {159, 187}

\bibitem[\protect\citeauthoryear{Plagemann, Kersting  \& Burgard}{Plagemann
  et~al.}{2008}]{Plagemann2008_nonStationaryGPs}
Plagemann C.,  Kersting K.,   Burgard W.,  2008, in Machine Learning and
  Knowledge Discovery in Databases. Springer Berlin Heidelberg, Berlin,
  Heidelberg, pp 204--219

\bibitem[\protect\citeauthoryear{Quadrianto, Kersting  \& Xu}{Quadrianto
  et~al.}{2010}]{Quadrianto2010_GPs}
Quadrianto N.,  Kersting K.,   Xu Z.,  2010, in Encyclopedia of Machine
  Learning. Springer US, Boston, MA, pp 428--439

\bibitem[\protect\citeauthoryear{Rajpaul}{Rajpaul}{2017}]{Rajpaul2017_thesis}
Rajpaul V.~M.,  2017, PhD thesis, University of Oxford

\bibitem[\protect\citeauthoryear{{Rajpaul}, {Aigrain}, {Osborne}, {Reece}  \&
  {Roberts}}{{Rajpaul} et~al.}{2015}]{Rajpaul2015_GPFramework}
{Rajpaul} V.,  {Aigrain} S.,  {Osborne} M.~A.,  {Reece} S.,   {Roberts} S.,
  2015, \mn@doi [\mnras] {10.1093/mnras/stv1428}, \href
  {https://ui.adsabs.harvard.edu/abs/2015MNRAS.452.2269R} {452, 2269}

\bibitem[\protect\citeauthoryear{Rasmussen \& Williams}{Rasmussen \&
  Williams}{2006}]{Rasmussen2006_GPs}
Rasmussen C.~E.,  Williams C. K.~I.,  2006, Gaussian Processes for Machine
  Learning.
{MIT Press}

\bibitem[\protect\citeauthoryear{{Robertson}, {Mahadevan}, {Endl}  \&
  {Roy}}{{Robertson} et~al.}{2014}]{Robertson2014_G581}
{Robertson} P.,  {Mahadevan} S.,  {Endl} M.,   {Roy} A.,  2014, \mn@doi
  [Science] {10.1126/science.1253253}, \href
  {https://ui.adsabs.harvard.edu/abs/2014Sci...345..440R} {345, 440}

\bibitem[\protect\citeauthoryear{{Saar} \& {Donahue}}{{Saar} \&
  {Donahue}}{1997}]{Saar1997}
{Saar} S.~H.,  {Donahue} R.~A.,  1997, \mn@doi [\apj] {10.1086/304392}, \href
  {https://ui.adsabs.harvard.edu/abs/1997ApJ...485..319S} {485, 319}

\bibitem[\protect\citeauthoryear{{Santos}, {Gomes da Silva}, {Lovis}  \&
  {Melo}}{{Santos} et~al.}{2010}]{Santos2010_StellarMagCycles}
{Santos} N.~C.,  {Gomes da Silva} J.,  {Lovis} C.,   {Melo} C.,  2010, \mn@doi
  [\aap] {10.1051/0004-6361/200913433}, \href
  {https://ui.adsabs.harvard.edu/abs/2010A&A...511A..54S} {511, A54}

\bibitem[\protect\citeauthoryear{{Santos} et~al.,}{{Santos}
  et~al.}{2014}]{Santos2014_HD41248}
{Santos} N.~C.,  et~al., 2014, \mn@doi [\aap] {10.1051/0004-6361/201423808},
  \href {https://ui.adsabs.harvard.edu/abs/2014A&A...566A..35S} {566, A35}

\bibitem[\protect\citeauthoryear{Schrijver \& Zwaan}{Schrijver \&
  Zwaan}{2000}]{SchrijverZwaan2000_pModes}
Schrijver C.~J.,  Zwaan C.,  2000, Solar and Stellar Magnetic Activity.
Cambridge Astrophysics, Cambridge University Press

\bibitem[\protect\citeauthoryear{Sokal}{Sokal}{1997}]{Sokal1997_ACT}
Sokal A.,  1997, in Functional Integration: Basics and Applications. Springer
  US, Boston, MA, pp 131--192

\bibitem[\protect\citeauthoryear{{Speagle}}{{Speagle}}{2020}]{Speagle2020_dynesty}
{Speagle} J.~S.,  2020, \mn@doi [\mnras] {10.1093/mnras/staa278}, \href
  {https://ui.adsabs.harvard.edu/abs/2020MNRAS.493.3132S} {493, 3132}

\bibitem[\protect\citeauthoryear{Stein}{Stein}{1999}]{Stein1999_MaternKernels}
Stein M.~L.,  1999, Interpolation of Spatial Data.
Springer New York

\bibitem[\protect\citeauthoryear{Styan}{Styan}{1973}]{Styan1973_hadamardProduct}
Styan G.~P.,  1973, \mn@doi [Linear Algebra and its Applications]
  {https://doi.org/10.1016/0024-3795(73)90023-2}, 6, 217

\bibitem[\protect\citeauthoryear{Wilcox}{Wilcox}{1972}]{Wilcox1972_SunSynodicPeriod}
Wilcox J.~M.,  1972, in Cosmic Plasma Physics. Springer US, Boston, MA, pp
  157--164

\bibitem[\protect\citeauthoryear{Wilson}{Wilson}{2014}]{Wilson2014_Thesis}
Wilson A.~G.,  2014, PhD thesis, University of Cambridge

\bibitem[\protect\citeauthoryear{Wilson, Knowles  \& Ghahramani}{Wilson
  et~al.}{2012}]{Wilson2012_GPRN}
Wilson A.~G.,  Knowles D.~A.,   Ghahramani Z.,  2012, in Proceedings of the
  29th International Coference on International Conference on Machine Learning.
  ICML'12.
Omnipress, USA, pp 1139--1146

\bibitem[\protect\citeauthoryear{{Zechmeister} \& {K{\"u}rster}}{{Zechmeister}
  \& {K{\"u}rster}}{2009}]{Zechmeister2009}
{Zechmeister} M.,  {K{\"u}rster} M.,  2009, \mn@doi [\aap]
  {10.1051/0004-6361:200811296}, \href
  {https://ui.adsabs.harvard.edu/abs/2009A&A...496..577Z} {496, 577}

\bibitem[\protect\citeauthoryear{{van Driel-Gesztelyi} \& {Green}}{{van
  Driel-Gesztelyi} \& {Green}}{2015}]{Driel2015_EvolutionOfActiveRegions}
{van Driel-Gesztelyi} L.,  {Green} L.~M.,  2015, \mn@doi [Living Reviews in
  Solar Physics] {10.1007/lrsp-2015-1}, \href
  {https://ui.adsabs.harvard.edu/abs/2015LRSP...12....1V} {12, 1}

\makeatother
\end{thebibliography}

\appendix
\section{Updates for variational means and covariances}
\label{app:variationalParams}

Closed-form updates for the variational parameters can be obtained from
mean-field theory \citep{Nguyen2013b_GPRN,Nguyen2015_thesis}. For the nodes, the
variational means and covariances are given by

\begin{equation}
\label{eq:sigmaF_muF}
\begin{aligned}
    \mu_{f_i}    &= \Sigma_{f_j} 
        \sum_{i=1}^{P} \sum_{j=1}^{Q} \frac{1}{\sigma_{y_{i}}^2}\left( 
            \bm{y}_i 
            - \sum\limits_{k \neq j}^{Q} \mu_{w_{ik}} \circ \mu_{f_{k}}  
        \right) \circ \mu_{w_{ij}} \\
    \Sigma_{f_j} &= \left[ 
        K_{f_j}^{-1} 
        + \sum_{i=1}^{P} \frac{1}{\sigma_{y_{i}}^2}\rm{diag}\Bigg(
            \mu_{w_{ij}} \circ \mu_{w_{ij}} 
            + \rm{diag} \left( \Sigma_{w_{ij}} \right) 
        \Bigg) \right]^{-1}
\end{aligned}
\end{equation}

\noindent
while for each weight they are given by

\begin{equation}
\label{eq:sigmaW_muW}
\begin{aligned}
    \mu_{w_{ij}}    &= \Sigma_{w_{ij}} 
        \sum_{i=1}^{P} \sum_{j=1}^{Q}  \frac{1}{\sigma_{y_{i}}^2} \left( 
            \bm{y}_i 
            - \sum\limits_{k \neq j}^{Q} \mu_{f_{k}} \circ \mu_{w_{ik}}  
        \right) \circ \mu_{f_{j}} \\
    \Sigma_{w_{ij}} &= \left[ 
        K_{w_{ij}}^{-1} 
        + \sum_{i=1}^{P} \frac{1}{\sigma_{y_{i}}^2} \rm{diag}\Bigg(
            \mu_{f_{j}} \circ \mu_{f_{j}} 
            + \rm{diag}\left( \Sigma_{f_{j}} \right) 
        \Bigg) \right]^{-1}.
\end{aligned}
\end{equation}
In these equations, $\sigma_i$ is calculated following
equation~\ref{eq:sigmaCalculation}, $\bm{y}_i$ is the vector containing the
measurements from dataset $i$, $K$ is the covariance matrix for the respective
node or weight, while diag and $\circ$ represent the matrix diagonal and the
Hadamard product \citep[e.g.][]{Styan1973_hadamardProduct}, respectively.

\section{Evidence lower bound}
\label{app:ELBO}
The ELBO of a GPRN model can be divided in three terms, as in equation
\eqref{eq:ELBO}. The first, known as the expected log-likelihood, is given by
\begin{equation}
    \label{eq:expectedLogLike}
    \begin{split}
        \mathbb{E}_q 
        [ \log &~p\left( D | \textbf{f}, \textbf{w}\right) ] = -\frac{1}{2}\sum_{n=1}^{N}\sum_{i=1}^{P} \left[ \log 2\pi \left(\sigma_{yerr_{ni}}^{2} + \sigma_{y_i}^{2} \right)\right]\\
        &~-\frac{1}{2\sigma^2}\sum_{n=1}^{N} \left(\mathcal{Y}_{n}^{T} - \mathcal{W}_{w_n}\mathcal{F}_{f_n} \right)^{T}\left(\mathcal{Y}_{n}^{T} - \mathcal{W}_{w_n}\mathcal{F}_{f_n} \right)\\
        &~-\frac{1}{2\sigma^2}\sum_{i=1}^{P}\sum_{j=1}^{Q}
        \left[ \text{diag} \left(\Sigma_{f_j}\right)^{T} \left(\mu_{w_{ij}}\circ \mu_{w_{ij}}\right)\right.\\
        &\left. +\;\text{diag} \left(\Sigma_{w_{ij}}\right)^{T} \left(\mu_{f_j}\circ \mu_{f_j}\right) +\text{tr}\left(\Sigma_{f_j}\Sigma_{w_{ij}}\right)\right],
\end{split}
\end{equation}

\noindent 
where tr represents the trace of the matrix, $\mathcal{Y}_{n}^{T}$ is a $1\times
P$ vector containing all observations at entry $n$, $\mathcal{W}_{w_n}$ is a
$P\times Q$ matrix containing the variational means $\mu_w$ at entry $n$, and
$\mathcal{F}_{f_n}$ is a $P\times 1$ vector containing the variational means
$\mu_f$ at entry $n$.

The second term, the expected log-prior, is 
\begin{equation}
\label{eq:expectedLogPrior}
\begin{split}
\mathbb{E}_q &~\left[ \log p(\textbf{f}, \textbf{w}) \right] = -\frac{1}{2}NQ\left(P+1\right)\log 2\pi\\
&~-\frac{1}{2}\sum_{j=1}^{Q}\left[\log |K_{f_j}| +\mu_{f_j}^{T}K_{f_j}^{-1}\mu_{f_j} +\text{tr}\left(K_{f_j}^{-1} \Sigma_{f_j} \right) \right]\\
&~-\frac{1}{2}\sum_{i=1}^{P}\sum_{j=1}^{Q}\left[\log |K_{w_{ij}}| +\mu_{w_{ij}}^{T}K_{w_{ij}}^{-1}\mu_{w_{ij}}+\text{tr}\left(K_{w_{ij}}^{-1} \Sigma_{w_{ij}} \right) \right],
\end{split}
\end{equation}

\noindent where $K$ is the covariance matrix of the respective node or weight. 

The last term is the entropy of the $q(\bm{f},\bm{w})$ distribution and can be
calculated with
\begin{equation}
\begin{split}
    \mathcal{H}_q [ q(\bm{f}, \bm{W}) ] = & ~ 
        \frac{1}{2} N Q (P+1) (1 + \log 2\pi) \\
        &~+ \frac{1}{2}\sum_{j=1}^{Q}\left[\log |\Sigma_{f_j}| \right] + \frac{1}{2}\sum_{i=1}^{P}\sum_{j=1}^{Q}\left[\log |\Sigma_{w_{ij}}|\right].
\end{split}
\end{equation}





\FloatBarrier
\section{Solar data analysis - posterior predictives}
\label{appendix:components_sun}

The posterior predictive means and standard deviations for the individual and
joint analyses of each time series, as described in section \ref{ch:Sun}, are
shown here. The full predictive means and its different components (mean, node
and weight functions) are shown separately in the left and right panels of the
following figures, respectively.

\begin{figure*}
    \centering
    \includegraphics[width=0.8\textwidth]{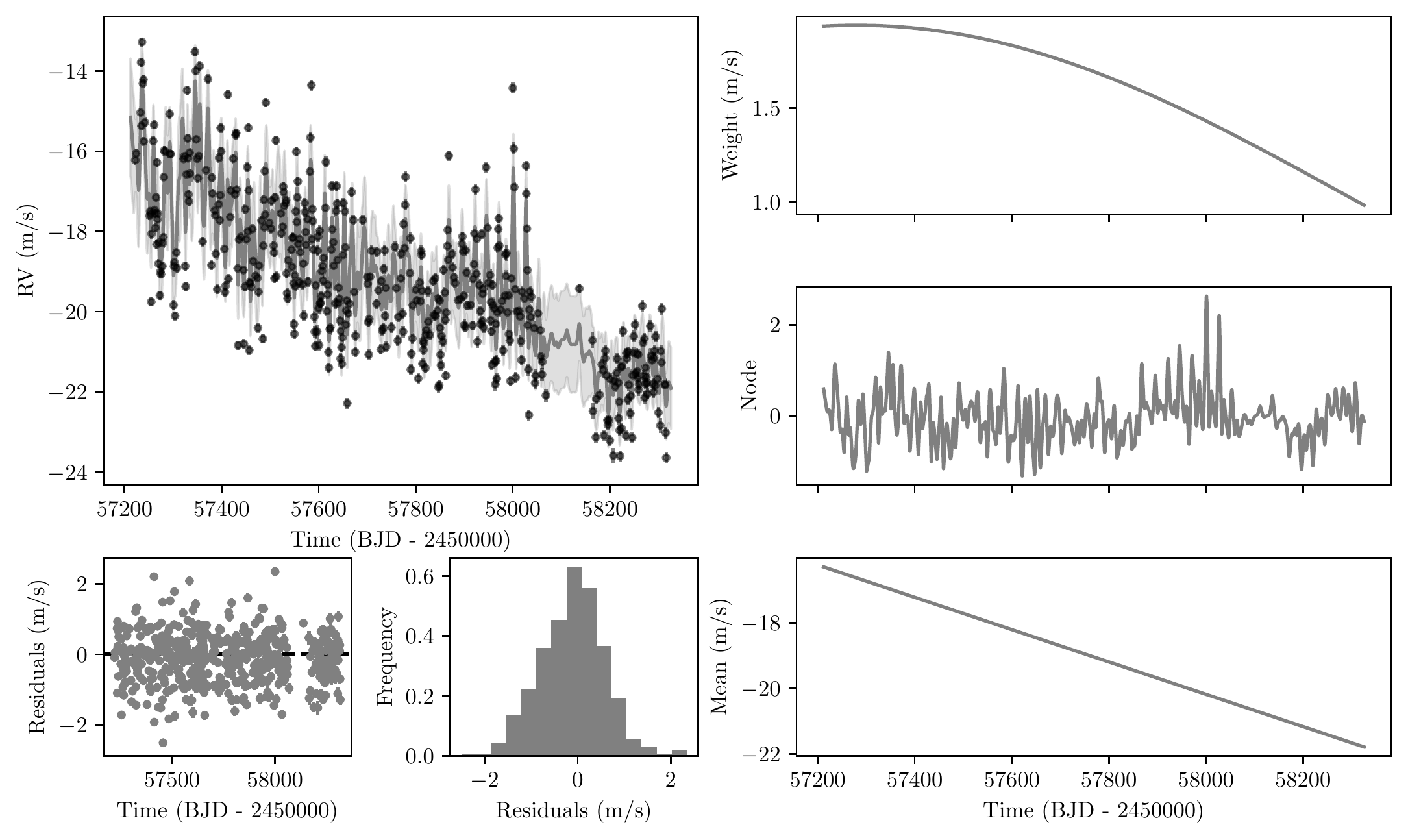}
    \caption{\textbf{On the top-left we show the RV data, as well as the 
    predictive means and standard deviations associated with the MAP values for 
    the GPRN hyper-parameters obtained when the RVs are analysed individually, while 
    on the bottom-left we show the residuals of the data with respect to the 
    predictive means and their histogram distribution.} On the right we plot, 
    from top to bottom, the weight, node, and mean functions that contribute to the 
    predictive means.}
    \label{fig:RV_fullPlots}
\end{figure*}

\begin{figure*}
    \centering
    \includegraphics[width=0.8\textwidth]{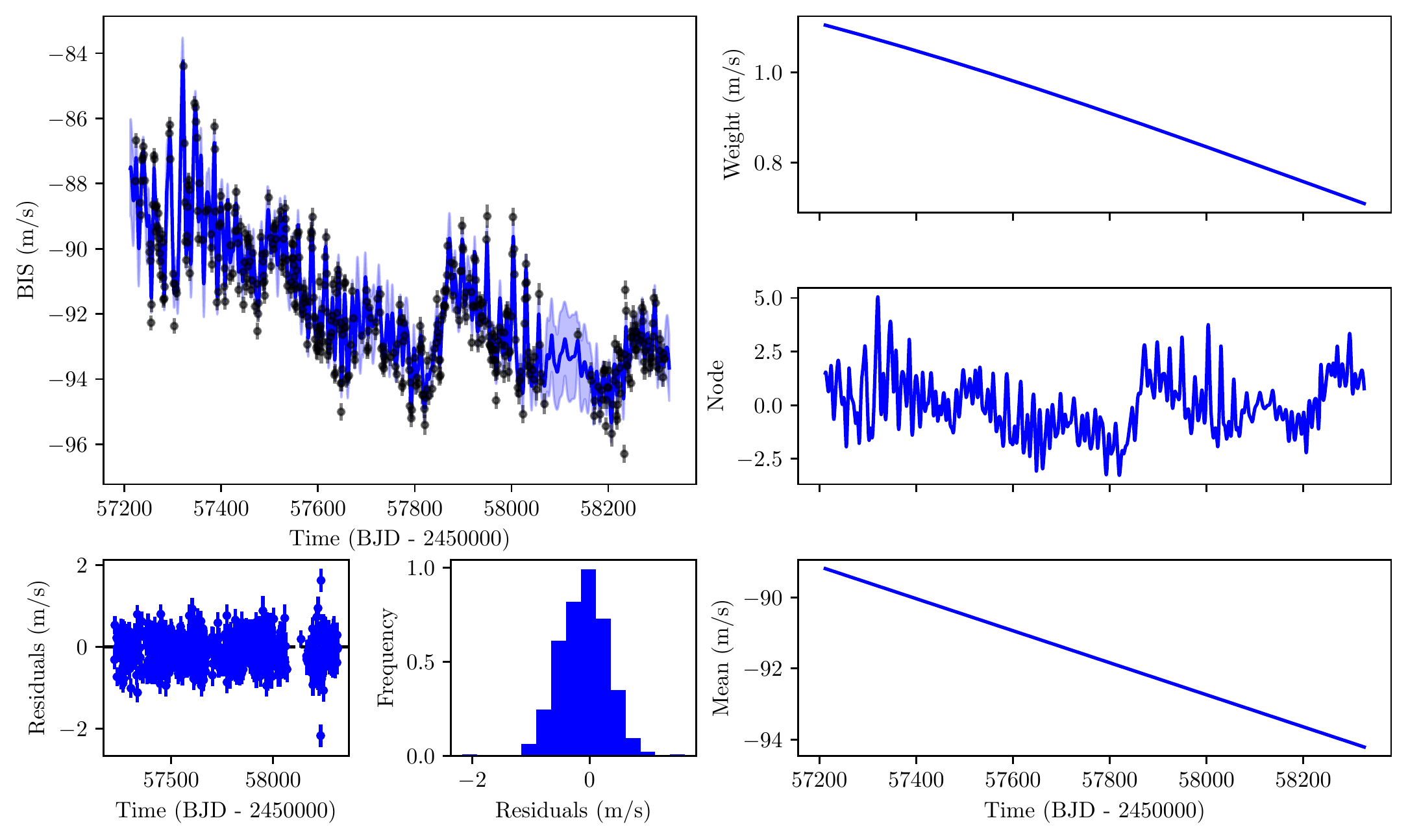}
    \caption{As in figure \ref{fig:RV_fullPlots}, but with respect to the
    individual analysis of the BIS data.}
    \label{fig:BIS_fullPlots}
\end{figure*}

\begin{figure*}
    \centering
    \includegraphics[width=0.8\textwidth]{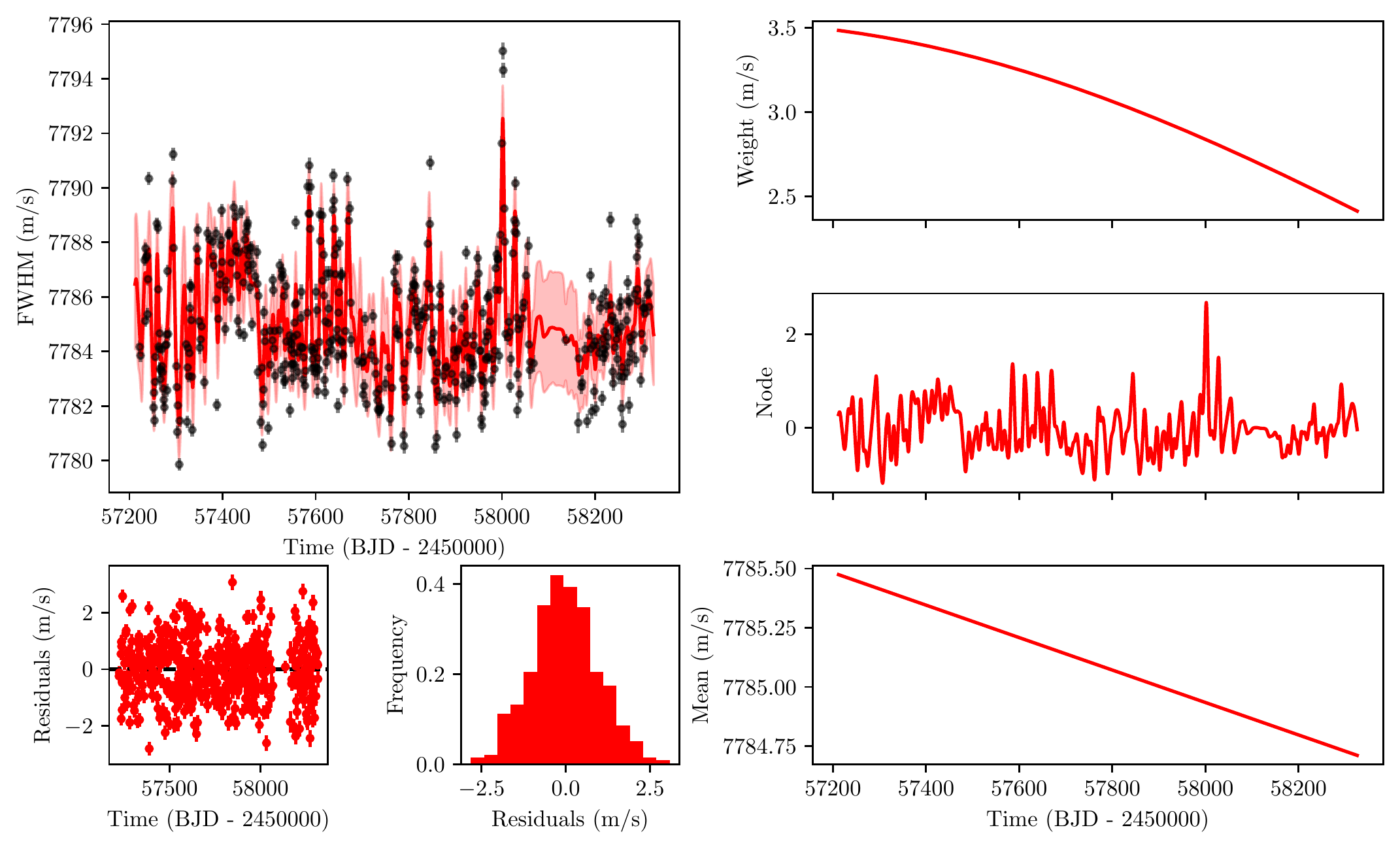}
    \caption{As in figure \ref{fig:RV_fullPlots}, but with respect to the
    individual analysis of the FWHM data.}
    \label{fig:FW_fullPlots}
\end{figure*}

\begin{figure*}
    \centering
    \includegraphics[width=0.8\textwidth]{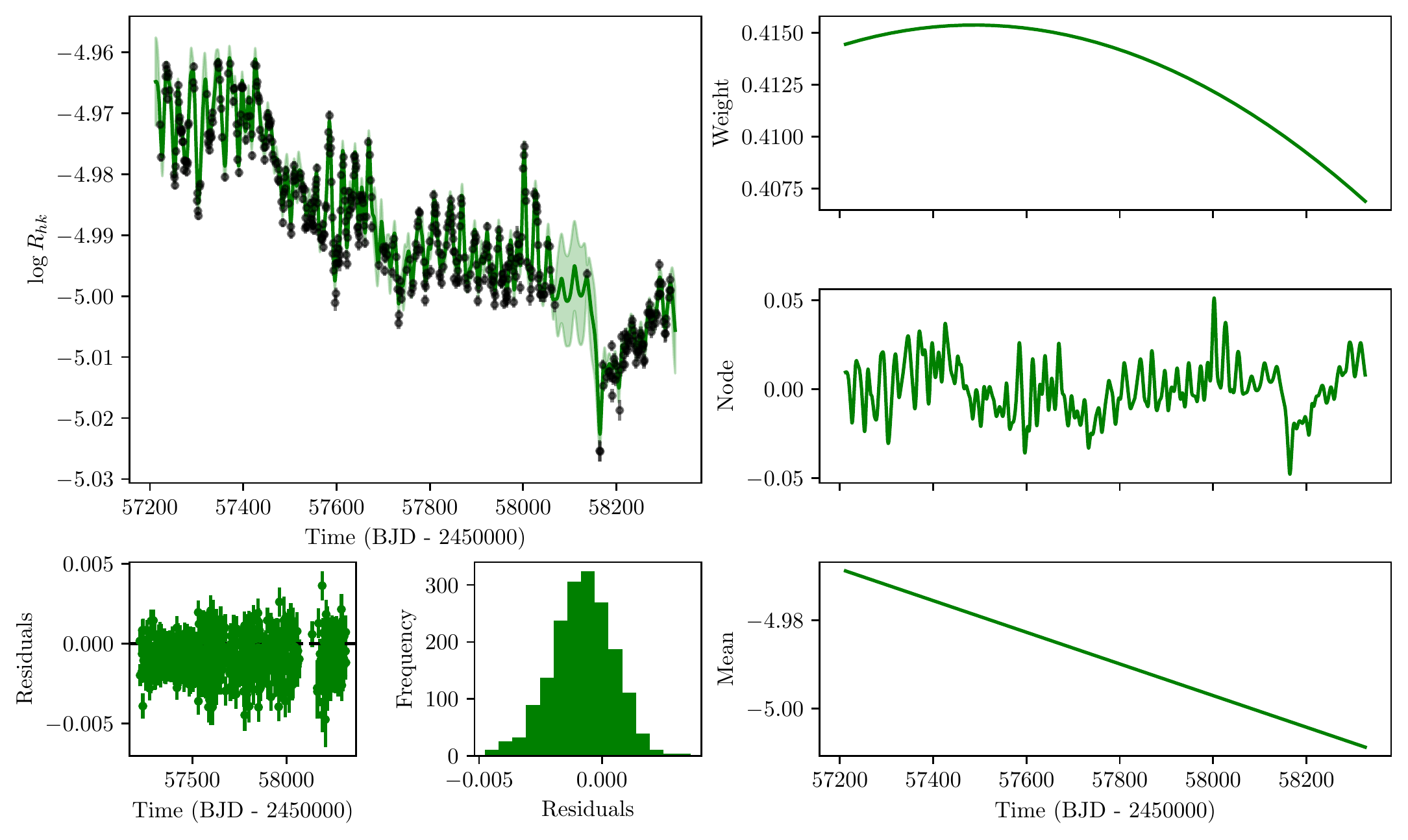}
    \caption{As in figure \ref{fig:RV_fullPlots}, but with respect to the
    individual analysis of the \logRhk data.}
    \label{fig:RHK_fullPlots}
\end{figure*}

\begin{figure*}
    \centering
    \includegraphics[width=0.8\textwidth]{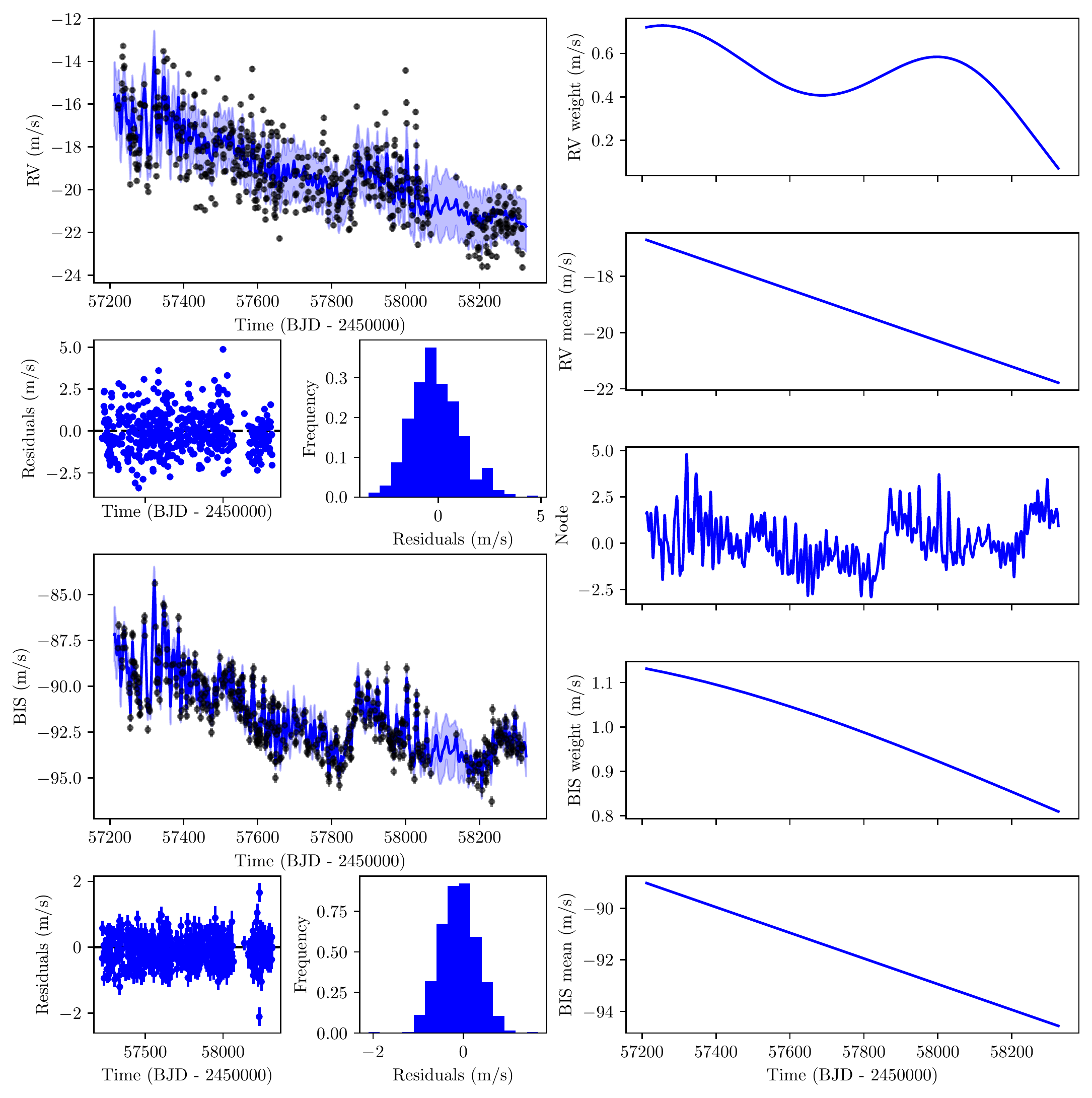}
    \caption{\textbf{On the left we show the RV (top) and BIS (bottom) data, as well 
    as the predictive means and standard deviations associated with the MAP values for 
    the GPRN hyper-parameters obtained when the RVs and BIS are analysed jointly. Below 
    each of those panels we show the respective residuals of the data with respect to the 
    predictive means and their histogram distribution.} On the right we show the contribution 
    to the predictive means due to the mean and weight functions for the RVs and BIS, as well 
    as the common node function in the middle.}
    \label{fig:fullPlots_RVBIS}
\end{figure*}

\begin{figure*}
    \centering
    \includegraphics[width=0.8\textwidth]{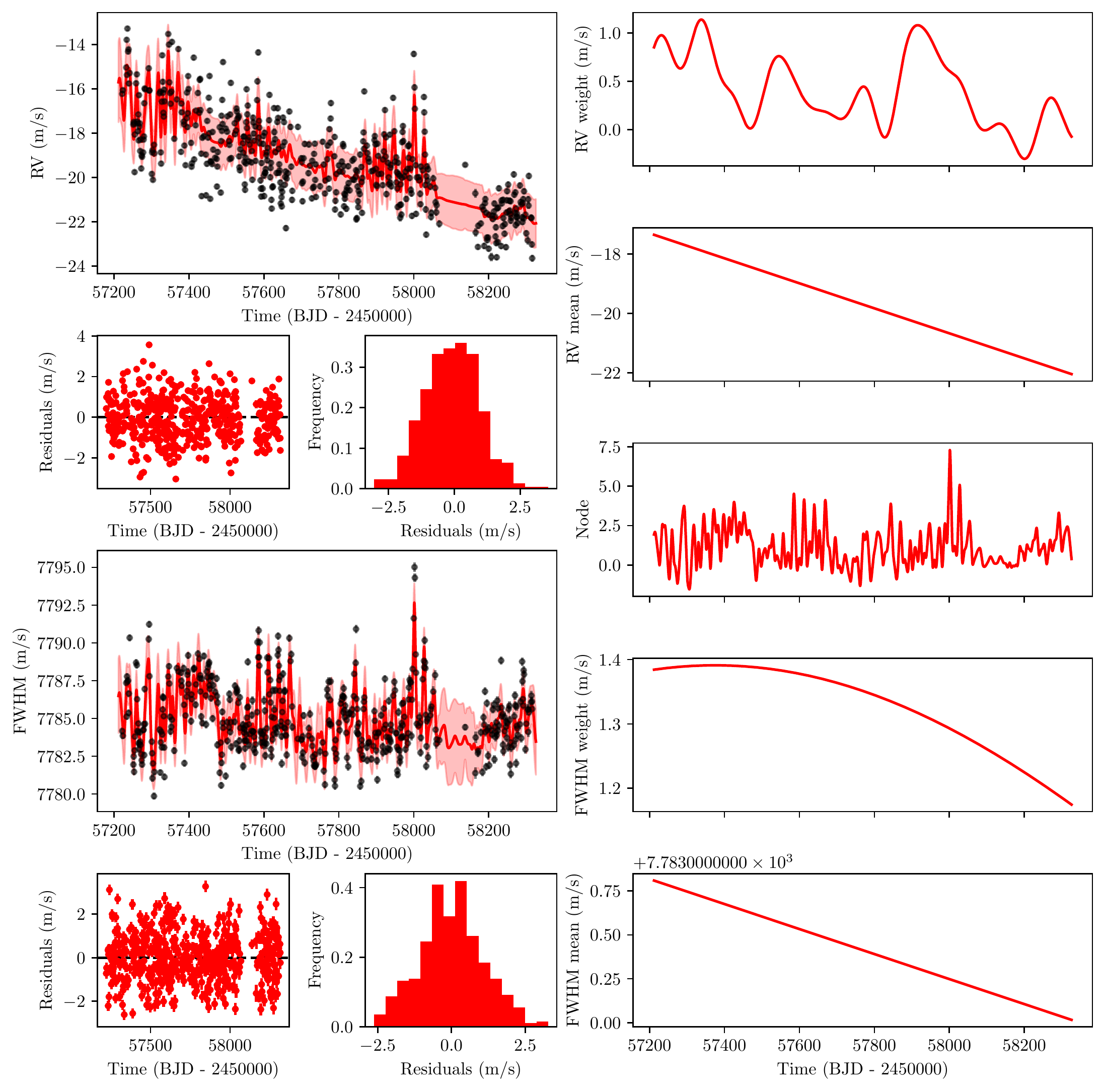}
    \caption{As in figure \ref{fig:fullPlots_RVBIS}, but with respect to the
    joint analysis of the RV and FWHM data.}
    \label{fig:fullPlots_RVFWHM}
\end{figure*}

\begin{figure*}
    \centering
    \includegraphics[width=0.8\textwidth]{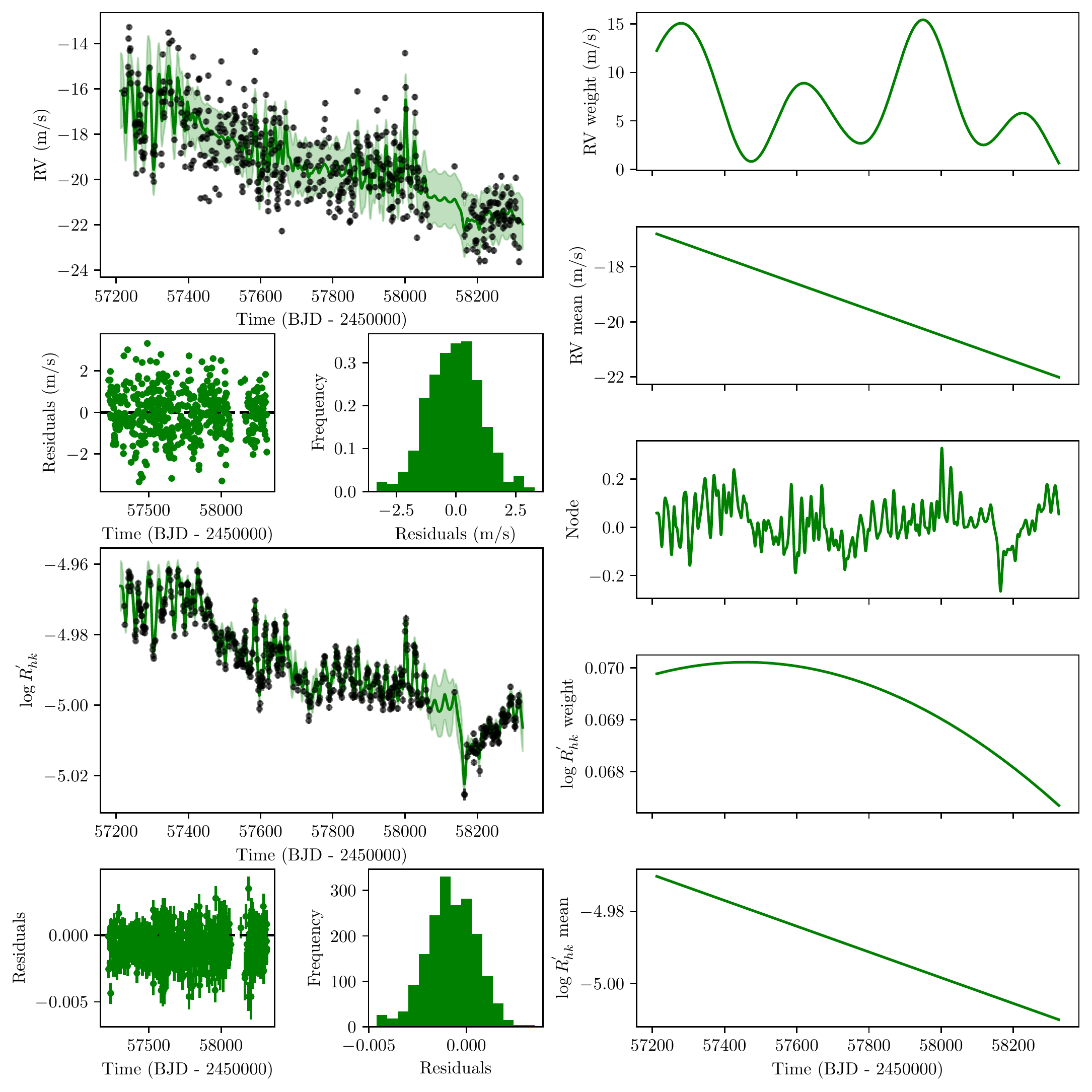}
    \caption{As in figure \ref{fig:fullPlots_RVBIS}, but with respect to the
    joint analysis of the RV and \textit{$\log R_{hk}^{'}$} data.}
    \label{fig:fullPlots_RVRHK}
\end{figure*}

\FloatBarrier

\section{Solar data analysis - marginal posterior distributions}
\label{appendix:corners_sun}

The corner plots with the marginal, single and pairwise, posterior distributions
obtained for all analyses of the HARPS-N solar data are shown here.

\begin{figure*}
    \centering
    \includegraphics[width=0.9\textwidth]{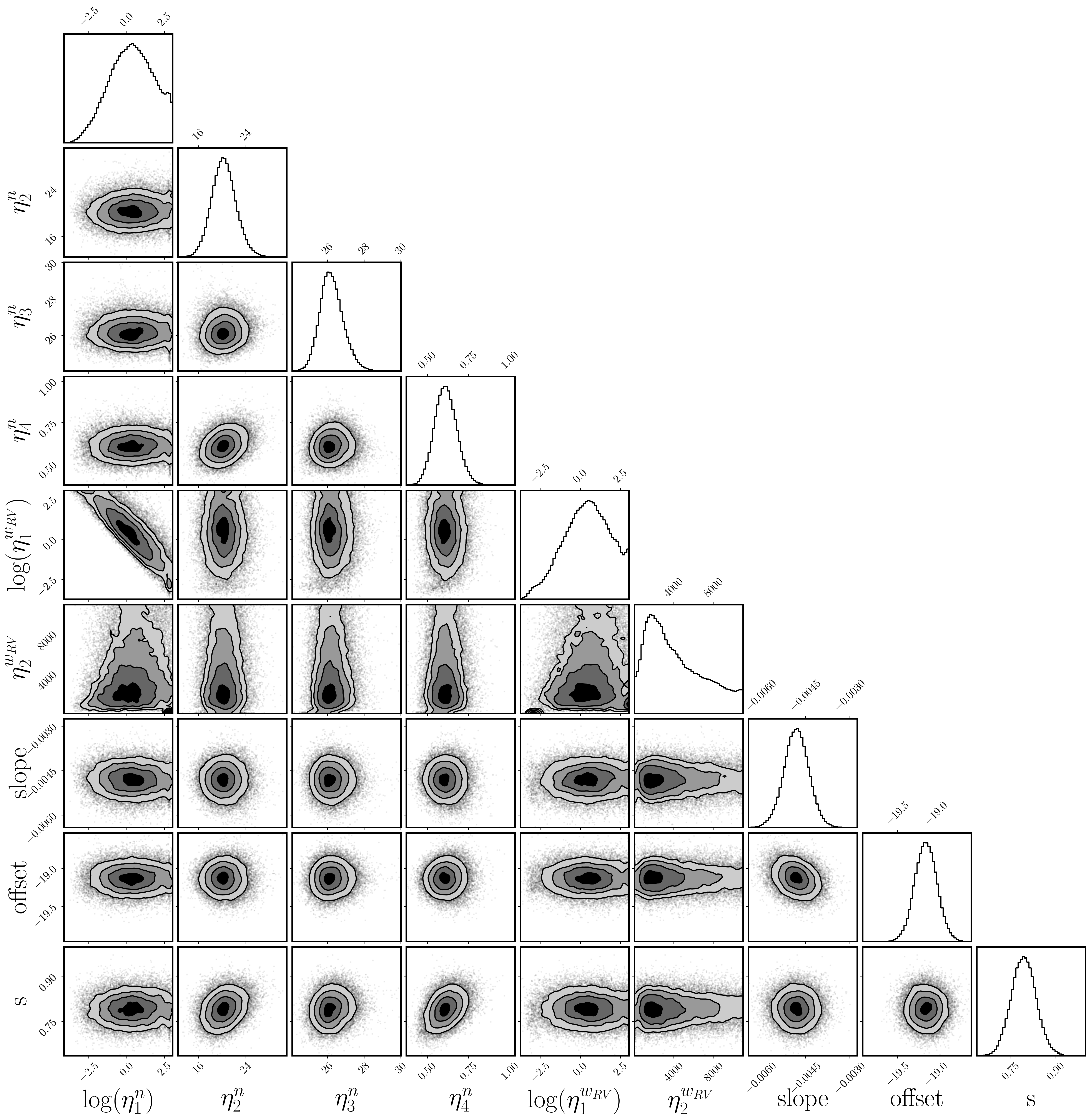}
    \caption{Corner plots showing the marginal, single and pairwise, posterior
    distributions for all GPRN hyper-parameters obtained when the RVs are
    analysed individually. In order to facilitate the visualization all
    $\eta_1$-type amplitudes are plotted in logarithmic form.}
    \label{fig:RV_gprn_corner}
\end{figure*}

\begin{figure*}
    \includegraphics[width=0.9\textwidth]{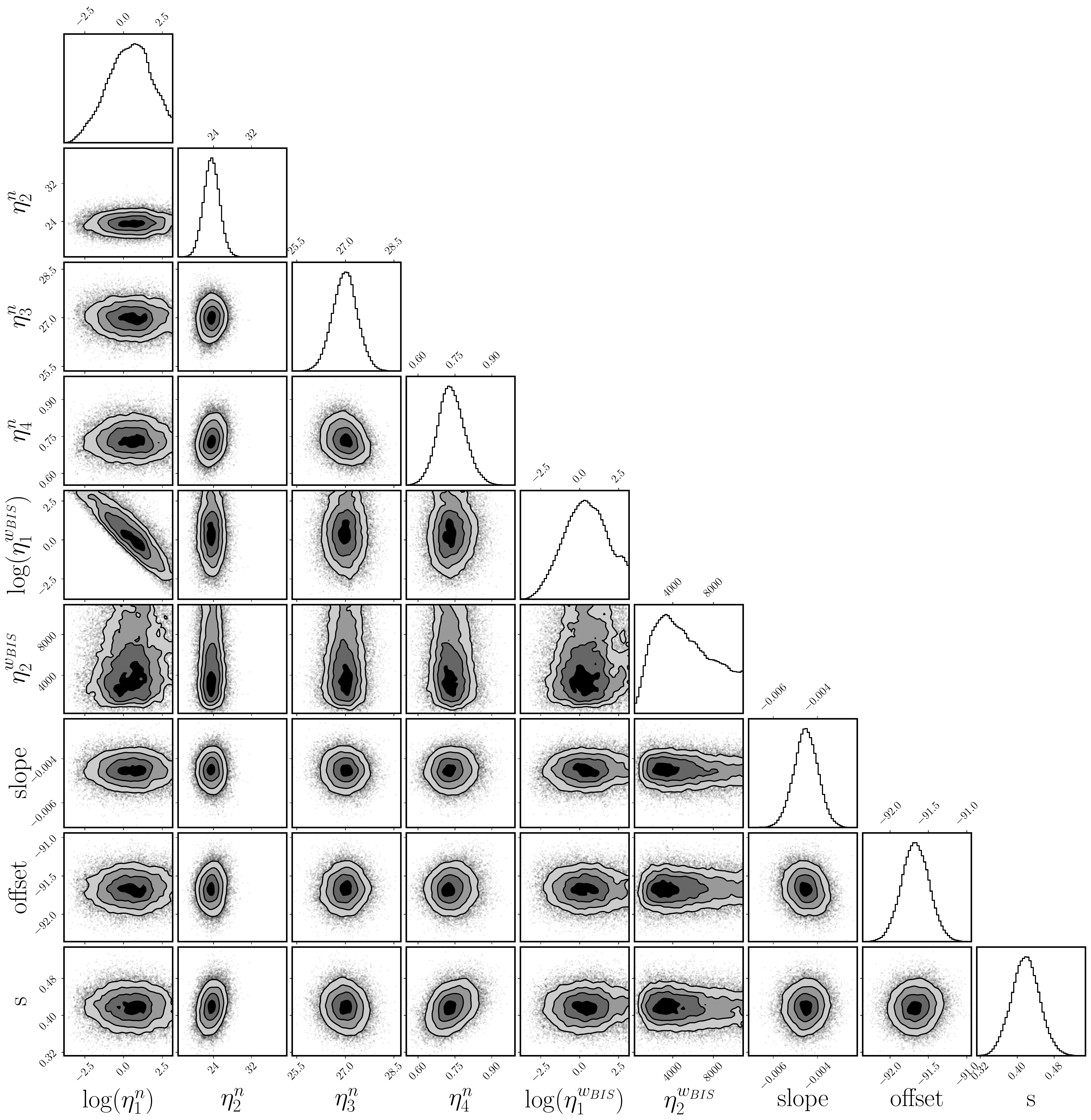}
    \caption{As in figure \ref{fig:RV_gprn_corner}, but with respect to the
    individual analysis of the BIS data.}
    \label{fig:BIS_gprn_corner}
\end{figure*}

\begin{figure*}
    \includegraphics[width=0.9\textwidth]{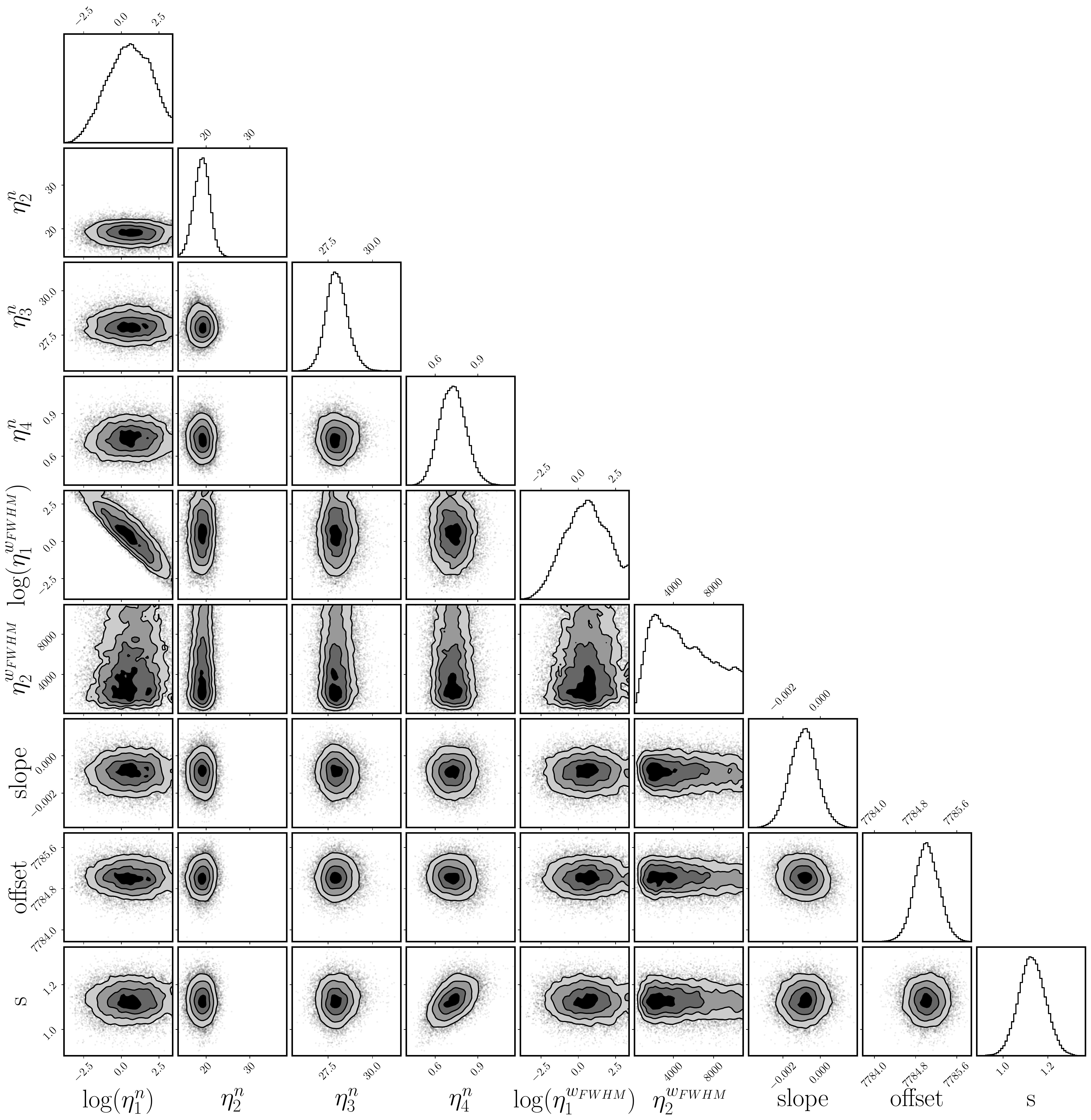}
    \caption{As in figure \ref{fig:RV_gprn_corner}, but with respect to the
    individual analysis of the FWHM data.}
    \label{fig:FWHM_gprn_corner}
\end{figure*}

\begin{figure*}
    \includegraphics[width=0.9\textwidth]{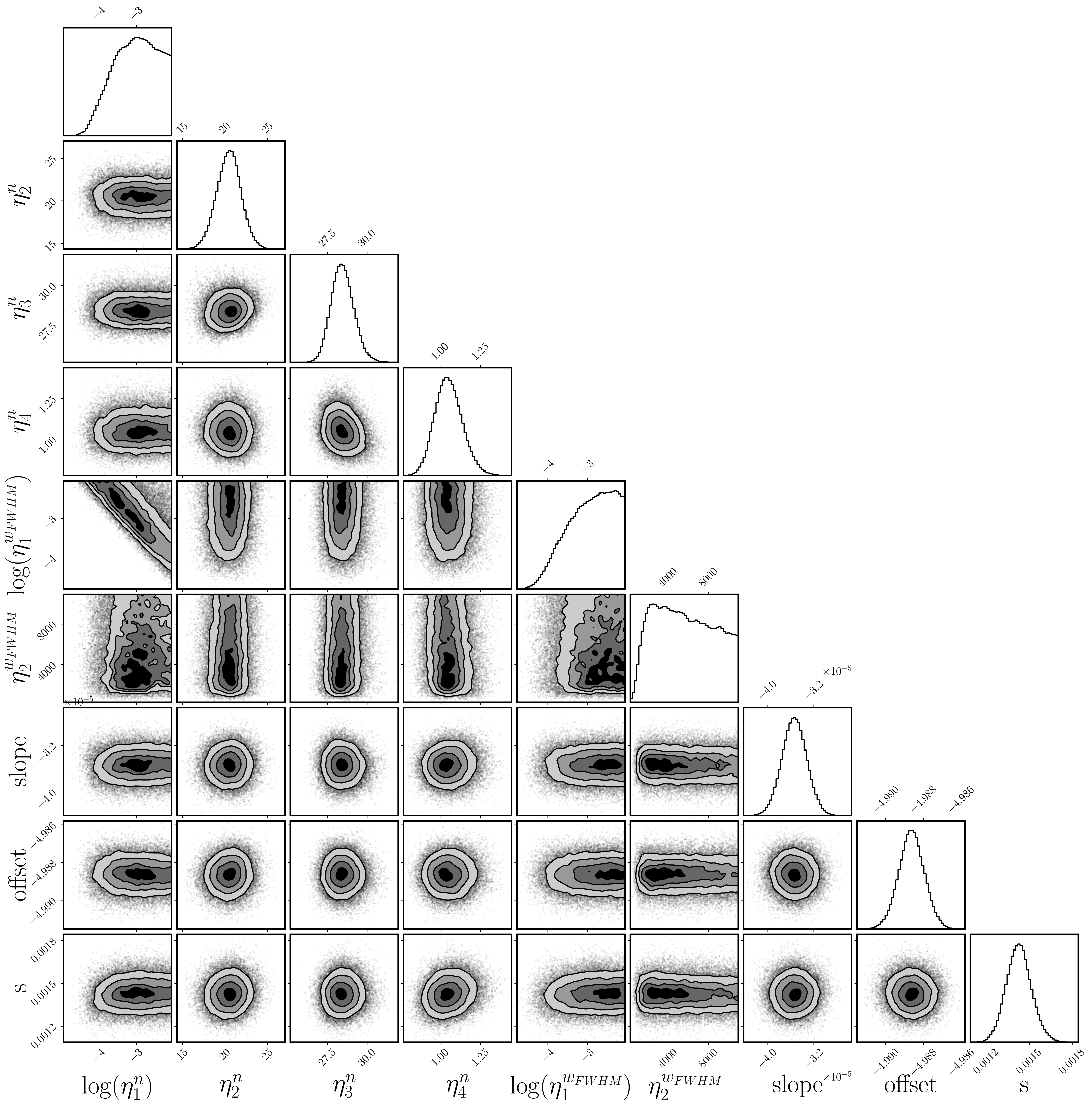}
    \caption{As in figure \ref{fig:RV_gprn_corner}, but with respect to the
             individual analysis of the \logRhk data.}
    \label{fig:Rhk_gprn_corner}
\end{figure*}

\begin{figure*}
    \begin{center}
    \includegraphics[width=0.9\textwidth]{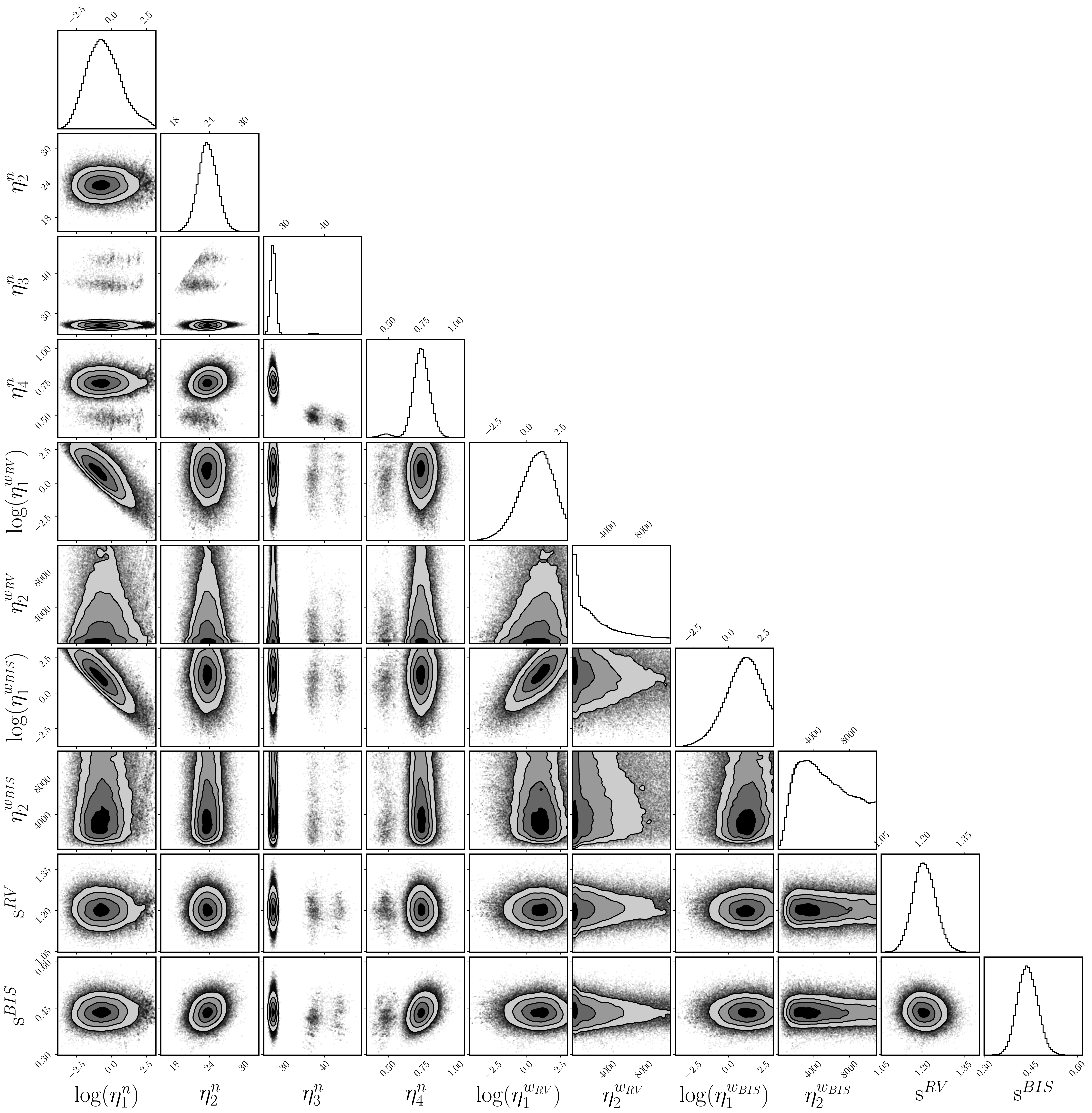}
    \end{center}
        \caption{Corner plots showing the marginal, single and pairwise,
        posterior distributions for all GPRN hyper-parameters obtained when the
        RVs and BIS are analysed jointly. In order to facilitate the
        visualization all $\eta_1$-type amplitudes are plotted in logarithmic
        form.}
    \label{fig:RVandBIS_gprn_corner}
\end{figure*}

\begin{figure*}
    \begin{center}
    \includegraphics[width=0.9\textwidth]{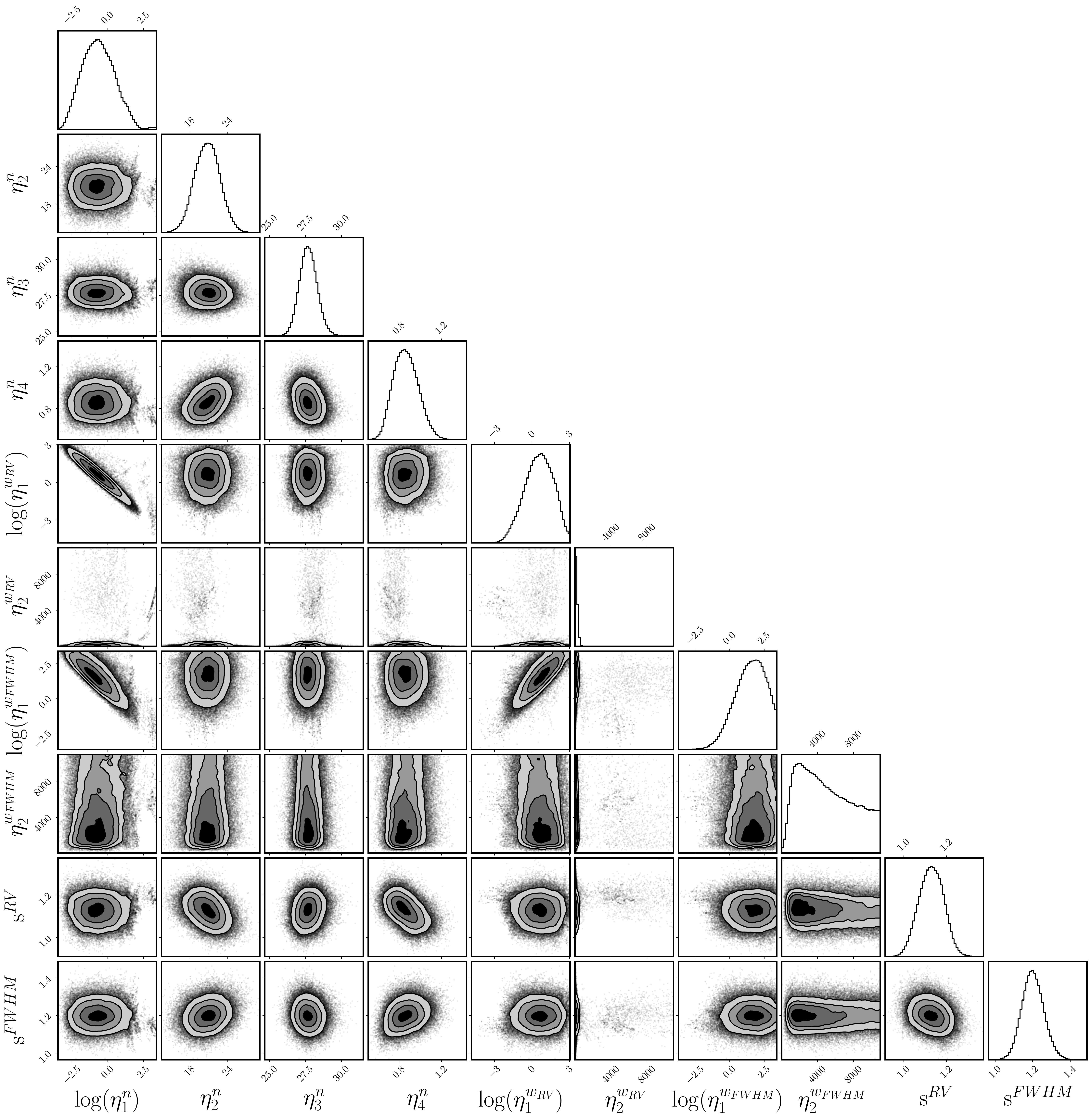}
    \end{center}
    \caption{As in figure \ref{fig:RVandBIS_gprn_corner}, but with respect to
    the joint analysis of the RV and FWHM data.}
    \label{fig:RVandFWHM_gprn_corner}
\end{figure*}

\begin{figure*}
    \begin{center}
    \includegraphics[width=0.9\textwidth]{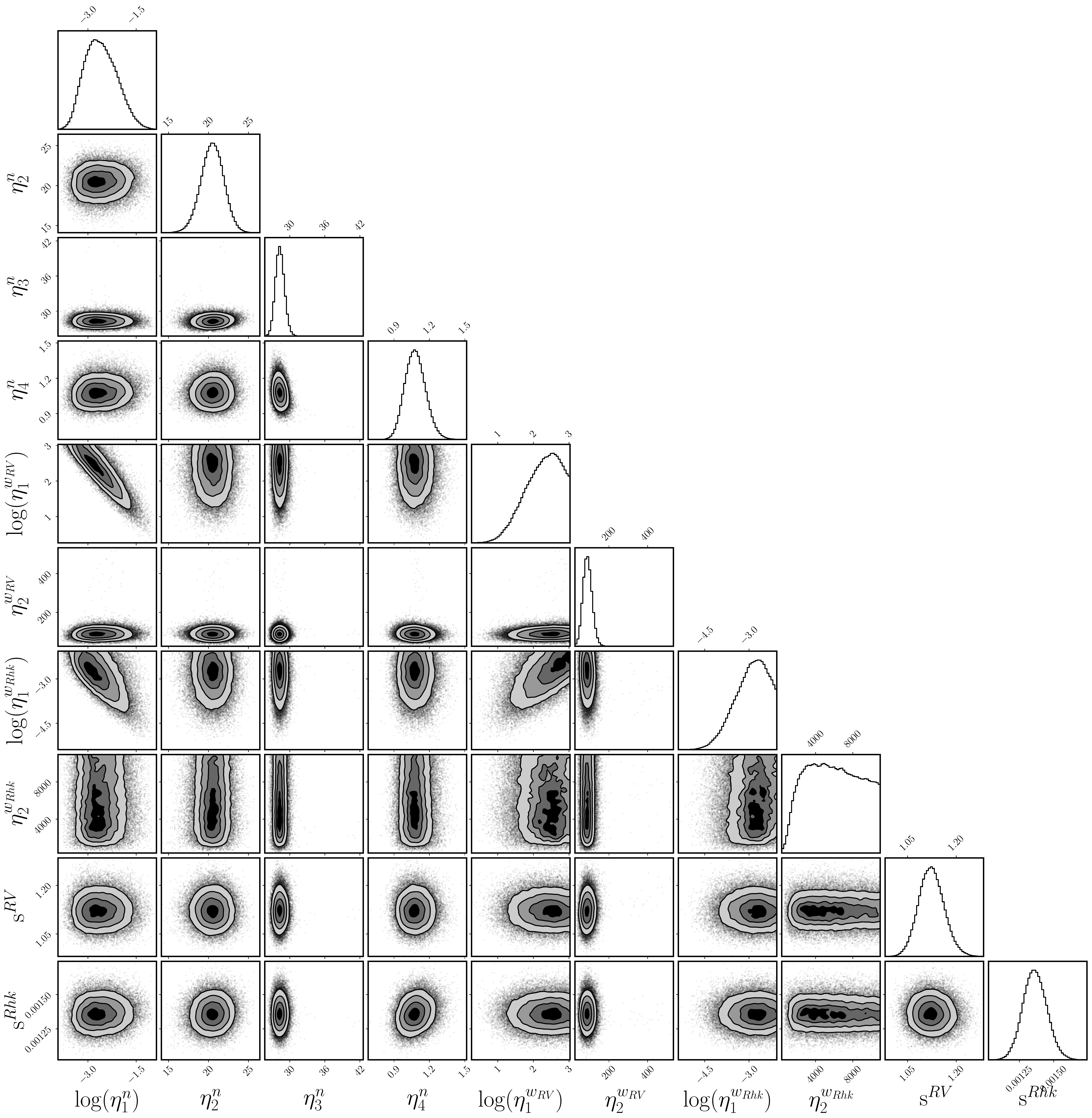}
    \end{center}
    \caption{As in figure \ref{fig:RVandBIS_gprn_corner}, but with respect to
    the joint analysis of the RV and \logRhk data.}
    \label{fig:RVandRhk_gprn_corner}
\end{figure*}

\FloatBarrier

\bsp	
\label{lastpage}
\end{document}